\documentclass[usegraphicx,usenatbib]{mn2e}

\def\Teff{$T_{\rm eff}$}
\def\logg{$\log\,g$}

\def\Vt{V${\rm t}$}
\def\Vr{V${\rm r}$}

\newcommand {\apgt} {\ {\raise-.5ex\hbox{$\buildrel>\over\sim$}}\ }
\newcommand {\aplt} {\ {\raise-.5ex\hbox{$\buildrel<\over\sim$}}\ }
 
\title[Neutron-capture elements in
open clusters]
{New insights on Ba over-abundance in open clusters.
\thanks{Based on observations collected at Paranal Observatory under program 088.D-0045}
Evidence for the intermediate neutron-capture process at play?}
\author[T.~Mishenina  et al.]
{T.~Mishenina$^{1,2}$,
M.~Pignatari$^{3,4}$,
G.~Carraro$^{5,6}$,
V.~Kovtyukh$^{1,2}$,
L.~Monaco${^5}$,
\newauthor
S.~Korotin$^{1,2}$,
E.~Shereta$^{1}$,
I.~Yegorova$^{5}$,
and F.~Herwig$^{4,7,8}$\\
$^{1}$Astronomical Observatory, Odessa National University,
       Shevchenko Park, 65014, Odessa, Ukraine\\
$^{2}$Isaac Newton Institute of Chile, Odessa branch,
       Shevchenko Park, 65014, Odessa, Ukraine\\
$^{3}$Department of Physics, University of Basel, Klingelbergstrabe 82,
        4056 Basel, Switzerland\\
$^{4}$NuGrid collaboration, http://www.nugridstars.org\\
$^{5}$European Southern Observatory, Alonso de Cordova 3107, 19001, Santiago de Chile, Chile\\
$^{6}$Dipartimento di Fisica e Astronomia, Universit\'a di Padova, Italy\\
$^{7}$Department of Physics \& Astronomy, University of Victoria, Victoria, BC, V8P5C2 Canada.\\
$^{8}$Joint Institute for Nuclear Astrophysics, USA.}

\begin{document}

\date{Accepted 2014 November 3. Received 2014 November 2; in original form 2014 August 17}
\pagerange{\pageref{firstpage}--\pageref{lastpage}}
\pubyear{2014}

\maketitle

\label{firstpage}

\begin{abstract}
Recently an increasing number of studies were devoted to measure the abundances
of neutron-capture elements heavier than iron in stars belonging to
Galactic Open Clusters (OCs). OCs span a sizeable range in  metallicity
(--0.6 $\leq$ [Fe/H] $\leq$ +0.4), and they show abundances of
light elements similar to disk stars of the same age. A different pattern
is  observed for heavy elements. A large scatter
is observed for Ba, with most OCs showing [Ba/Fe] and [Ba/La] overabundant
with respect to the Sun.
The origin of this overabundance is not clearly understood.
With the goal of providing new observational insights 
we determined radial velocities, atmospheric parameters and chemical
composition of 27 giant stars members of five
OCs: Cr~110, Cr~261, NGC~2477, NGC~2506 and NGC~5822.
We used high-resolution spectra obtained with the UVES spectrograph
at ESO Paranal. We perform a detailed
spectroscopic analysis of these stars to measure the abundance of up
to 22 elements per star.
We study the dependence of element abundance on metallicity and age
with unprecedented detail, complementing our analysis with data culled from 
the literature. We confirm the trend of Ba overabundance
in OCs, and show its large dispersion for clusters younger than
$\sim$4 Gyr. Finally, the implications of our results for stellar
nucleosynthesis are discussed. 
We show in this work that the Ba enrichment compared to other neutron-capture
elements in OCs 
cannot be explained by the contributions from the slow neutron-capture
process and the rapid neutron-capture process. Instead, we argue that
this anomalous signature can be explained by assuming an additional
contribution by the intermediate neutron-capture process.
\end{abstract}


\section{Introduction}
In \cite{mis13b} we reported on the detailed
chemical abundance analysis of giant stars in the open clusters (OCs)
Ruprecht~4, Ruprecht~7, Berkeley~25, Berkeley~73, Berkeley~75, NGC~6192,
NGC~6404, and NGC~6583.
Our analysis was focused on neutron-capture elements located at the first
and second
neutron-magic peaks beyond iron (N=50 and 82, respectively).
In the Solar System about half of the abundance beyond Fe are made by the
slow neutron capture process \citep[$s$-process, e.g.,][and references
therein]{kaeppeler:11}, while the other half is made by the rapid
neutron capture process
\citep[$r$-process, e.g.,][and references therein]{thielemann:11}.
On average, most OCs have a metallicity around the Sun (with some exceptions), therefore
any relevant departure from solar abundances of heavy elements
provides important insights about OCs formation and about the
production of these elements in stars.
Using as a reference the Solar System, heavy elements that are
mostly produced by the $s$-process are usually called $s$-process
elements. Ba and La are typical examples of this group, located at
the neutron shell closure N=82. According to the residual method,
heavy elements that instead are not produced efficiently by the $s$-process are $r$-process elements, e.g., Eu \citep[e.g.,][]{bisterzo:14}.
Galactic chemical evolution simulations have shown that starting from
Ba and for heavier elements the residual method provides results that
are quite consistent with spectroscopic observations of old metal-poor
$r$-process-rich stars \citep[][]{travaglio:04}.
In this work we will use the same naming scheme of $s$-process and $r$-process
elements for OCs.
Between the Sr neutron-magic peak and Xe, the residual method seems to
fail to reproduce the solar system inventory, requiring the introduction
of an alternative nucleosynthesis component, called
Lighter Element Primary Process, or LEPP \citep[][]{travaglio:04}.
If this component is the same observed in a sample of old metal-poor
stars in the galactic halo is still matter of debate \citep[][]{montes:07}.
A larger amount of stellar data are becoming available in the last years
for metal-poor stars, including abundances of elements in the mass region
between Sr and Ba, e.g., Ag and Pd \citep[e.g.,][]{hansen:12}.
This will allow in the near future to better constrain the origin
of the LEPP at low metallicity.
Different nucleosynthesis processes have been proposed as a source
of the LEPP, in the early Galaxy and eventually in the Solar System
\citep[][]{hoffman:96,frohlich:06,qian:08,pignatari:08,farouqi:09,arcones:11,frischknecht:12}.
Recently, the existence of the LEPP for the solar system has been
questioned, and observations of heavy elements in OCs compared to
the Sun were one of the main arguments used to support this
analysis \citep[][]{maiorca:12,trippella:14}.

Nevertheless, the peculiar high Ba abundance compared to Fe and
other heavy elements with respect to the Sun observed in a number of OCs,
remains a puzzle.
From available data, Ba overabundance seems to be present at any age
and metallicity, and seems to increase at decreasing age \citep[][]{mai11,dor12,
yon12,jfr13,mis13b}. The origin of this overabundance, however, is not
understood, and the data analysis far from being  homogeneous.
One way to get more insight on this problem is to study the overabundance
in a wider age and metal abundance range.
To this aim, in this study we add to the original Mishenina et al. (2013b) sample five more  open clusters:
Cr 110, Cr 261, NGC 2477, NGC 2506, and NGC 5822, allowing to cover within a
consistent analysis a wider range in metallicity ($-0.2 \leq [Fe/H] \leq +0.15 $) and age
(0.5 to 7.0 Gyr).

Previous studies are available for all these clusters, with partial overlap.
In particular, three stars in Cr 110 were studied by~\cite{pan10}, 6 stars
in Cr 261 from \cite{car05},  6 star in NGC~2477 from \cite{bra08},
4 stars in NGC~2506 from \cite{car05}, and, lastly, 3 stars in NGC~5822
from \cite{san09}.  We anticipate that a good agreement is in general
obtained for all the stars in common.
Some exceptions are present for Ba, where we found discrepancies up
to 0.3 dex between different works for the [Ba/Fe], and 0.4 dex
in one case.

Additionally, the same observational material presented here for NGC~2477
and NGC~5822,
has also been recently analyzed by \citep[hereafter C14,][]{caffau14}.
The C14 study allows for an independent cross check on 
the atmospheric parameters and iron content derived for the program
stars and provide an assessment on the typical differences
on these parameters as derived by different researchers even when
adopting similar, although not identical techniques.

The data quality and origin, and the analysis techniques are
identical to \cite{mis13a}. In particular, non-local thermodynamic
equilibrium NLTE conditions are adopted in deriving Ba abundance.

The layout of the paper is as follows.
In Section~2 we describe how data were collected and reduced.
Section~3 is devoted to the determination of the stars' photospheric
parameters (effective temperature $T_{\rm eff}$,  surface gravity \logg, and micro-turbulence velocity \Vt),
while Section~4 illustrates how we perform the abundance analysis.
Our results, together with a comparison with literature material,
are discussed in Section~5. Finally, in Section~6 we discuss
our results in the framework of stellar nucleosynthesis.
Conclusions and final remarks are given in Section~7.

\section{Observations and data reduction}

The main parameters: galactic coordinates (for J2000.0), galactocentric distance
$R_{GC}$ and age,  of the investigated clusters are listed in the
Table \ref{par}, together
with the observation epochs and signal-to-noise (SNR) range.
Age and distances are obtained from the sources listed in the last column.
In particular, Galacto-centric distances  have been re-scaled
to a Sun distance to the Galactic center of 8.5 kpc.

\begin{table*}
\caption{The main parameters of the investigated clusters. The last column indicates the source for age and distance.}
\label{par}
\begin{tabular}{lcrrlrrcl}
\hline
Name       &     l   &     b   &$R_{\small GC} $ &    age  & Exposure& Date & SNR & \\
           &    deg  &    deg  & kpc &    Gyr  &   sec &   &   &\\
\hline
Collinder 110  &  209.649&--01.927&    10.2   & 1.3  & 2$\times$2000 & Feb 28, Mar 06 (2012) & 26$-$64      & Bragaglia \& Tosi (2003)\\
Collinder 261  &  301.684&--05.528&     7.5   & 7.0  &  3$\times$2400 & Feb 24, Mar 01, 06 (2012) & 39$-$53 & Gozzoli et al. (1996)\\
NGC 2477       &  253.563&--05.838&     8.9   & 0.6  &  3$\times$1500 & Oct 28 (2011), Mar 08 (2012) &  66$-$92 & D'Orazi et al. (2009)\\
NGC 2506       &  230.564&  09.935&    10.9   & 1.9  & 2$\times$2000 & Feb 03, Mar 07 (2012) & 20$-$87 & Reddy et al. (2012)\\
NGC 5822       &  321.577&  03.585&     7.9   & 0.45 &  3$\times$1000& Mar 01, 06, 24 (2012) & 92$-$108  & Carraro et al. (2011)\\
\hline
\end{tabular}
\end{table*}

\begin{table*}
\caption{The main parameters of the investigated stars.}
\label{par2}
\begin{tabular}{lcccccclrcc}
\hline
Name  &    RA(2000.0) & Dec(2000.0) &    $V$    &    $B-V$  &  \Teff       &\logg  &    \Vt    &    [Fe/H] & \Vr &Membership \\
\hline
    & deg & deg & mag & mag & $^oK$ & & km s$^{-1}$ & & km s$^{-1}$ \\
\hline
Cr 110  &           &        &             &          &       &          &           &    &           \\
1122    &   99.705000 &  2.108611   &13.740  &   1.383  &   4954      &  2.6  &   1.2    &  --0.05   &  38.19$\pm$0.10  &     M      \\
1134    &   99.687500 &  2.073194 & 13.704  &   1.360  &   4940      &  2.6  &   1.2    &    0.02   &  38.14$\pm$0.13  &     M    \\
1149    &   99.712917 &  2.065083 & 13.637  &   1.389  &   4906      &  2.6  &   1.2    &  --0.01   &  37.46$\pm$0.39  &     M     \\
1151    &   99.726667 &  2.066278  & 13.691  &   1.327  &   4956      &  2.6  &   1.2    &    0.02   &  37.94$\pm$0.04  &     M    \\
2129    &   99.671250 &  2.018139  & 13.656  &   1.340  &   4933      &  2.6  &   1.2    &  --0.04   &  38.69$\pm$0.11  &     M     \\
3122    &   99.644583 &  2.028056 & 13.464  &   1.378  &   4758      &  2.4  &   1.0    &  --0.03   &  39.94$\pm$0.05  &     M     \\
            &             &          &             &       &          &           &    &      &     \\
Cr 261  &             &          &             &       &          &           &    &       &    \\
2269    &   189.412917 &--68.386806  & 14.241  &   1.403  &   4575      &  2.4  &   1.2    &  --0.02   &--28.03$\pm$0.14  &  M         \\
2291    &   189.480417 &--68.413861  & 13.572  &   1.328  &   4746      &  2.5  &   1.2    &    0.00   &--24.18$\pm$0.14 &  M        \\
2309    &   189.551667 &--68.342139   & 13.718  &   1.286  &   4746      &  2.5  &   1.2    &    0.00   &--26.23$\pm$0.16 &  M         \\
2311    &   189.545000 &--68.392778 & 14.164  &   1.362  &   4778      &  2.5  &   1.15   &  --0.02   &--25.56$\pm$0.15 &   M       \\
2313    &   189.556667 &--68.399333  & 14.011  &   1.448  &   4674      &  2.5  &   1.2    &  --0.01   &--23.20$\pm$0.11 &    M      \\
                 &             &          &             &       &          &           &    &       &    \\
NGC 2477&             &          &             &       &          &           &    &        &   \\
4027    &  118.087917 &--38.577194   & 12.153  &   1.198  &   4966      &  2.7  &   1.4    &    0.10   &   7.03$\pm$0.13   &  M         \\
4221    &  118.152083 &--38.631750 & 12.270  &   1.171  &   4975      &  2.8  &   1.2    &    0.19   &   8.80$\pm$0.23    &  M         \\
5043    &  118.040417 &--38.598306   & 12.165  &   1.170  &   5001      &  2.8  &   1.2    &    0.08   &   13.22$\pm$0.27 &  NM         \\
5076    &  118.061667 &--38.629194   & 12.410  &   1.220  &   4954      &  2.7  &   1.2    &    0.18   &   9.22$\pm$0.33 &  M         \\
7266    &  117.955000 &--38.535694    & 12.252  &   1.193  &   4966      &  2.8  &   1.2    &    0.19   &   9.30$\pm$0.14 &  M         \\
7273    &  117.947917 &--38.543389	    & 12.390  &   1.174  &   4985      &  2.8  &   1.2    &    0.20   &   8.77$\pm$0.51 &  M         \\
8216    &  118.064583 &--38.457306   & 12.334  &   1.272  &   4945      &  2.7  &   1.2    &    0.14   &   3.99$\pm$0.50 &  NM        \\
                 &             &          &             &       &          &           &    &      &     \\
NGC 2506&             &          &             &       &          &           &    &       &    \\
1112    &  120.013750 &--10.762250   & 12.961  &   0.958  &   4969      &  2.6  &   1.2    &  --0.22   &  83.99$\pm$0.27 &     M     \\
1229    &  120.030833 &--10.740722  & 13.118  &   1.011  &   4728      &  2.4  &   1.0    &  --0.22   &   82.54$\pm$0.58 &       M   \\
2109    &  120.029583 &--10.779000   & 13.146  &   0.890  &   5040      &  2.6  &   0.9    &  --0.22   &  89.31$\pm$0.05  &     NM     \\
2380    &  120.038750 &--10.818806  & 13.187  &   0.927  &   4992      &  2.6  &   1.0    &  --0.19   &  83.64$\pm$0.53 &      M    \\
3231    &  119.982917 &--10.805944  & 13.105  &   0.952  &   4974      &  2.6  &   1.2    &  --0.22   &  84.36$\pm$0.51  &       M   \\
5271    &  120.028750 &--10.752000 & 13.204  &   0.923  &   4993      &  2.6  &   1.15   &  --0.24   &  83.52$\pm$0.15  &      M     \\
                 &             &          &             &       &          &           &    &       &    \\
NGC 5822&             &          &             &       &          &           &    &        &   \\
13292   &   226.164167 &--54.351139    &  10.401 &  1.040    &   5010      &  2.8  &   1.2    &   0.04    & --29.35$\pm$0.34  &      M     \\
16450   &   226.059167 &--54.429833    &  10.281 &  1.050   &   4972      &  2.6  &   1.2    & --0.02    & --25.69$\pm$0.37  &       NM    \\
18897   &   225.955833 &--54.336278    &  10.842 &  1.014    &   5030      &  2.7  &   1.0    & --0.02    & --29.01$\pm$0.22  &       M    \\
 2397   &   226.071250 &--54.473111     &  10.455 & 1.010      &   5036      &  2.8  &   1.1    &   0.02    & --29.67$\pm$0.79&       M    \\
\hline
\end{tabular}
  \\
Notes. The data of $V$ and $B-V$ were taken from \cite{bra03}
for  Cr~110,  from \cite{goz96}  for Cr~261,  from
\cite{kas97} for NGC~2477, from \cite{mar97}
for NGC~2506, and  from \cite{carr11} for NGC~5822.
 \end{table*}

Observations were taken in service mode using the multi-object fibre-fed
FLAMES facility mounted at the
ESO-VLT/UT2 telescope at the Paranal Observatory (Chile).  Two
or three exposures (depending on the cluster, see Table~1)
were taken with  the red arm of the UVES high-resolution spectrograph. 
The UVES spectrograph was
set up around a 5800\AA\, central wavelength, thus covering the 4760--6840\AA\,
wavelength range and providing a resolution of R$\simeq$47,000.

Radial velocities (see Table~2) were computed using the IRAF/{\tt fxcor}
task to
cross-correlate the observed spectra with a synthetic one from the
\cite{coelho:05} library with stellar parameters  \Teff=5250\,K,
log\,$g$=2.5, solar
metallicity, and no $\alpha$-enhancement. The IRAF {\tt rvcorrect} task was
used to calculate the correction from geocentric velocities to heliocentric.

We took the
stars radial velocity to be the average of the two/three epochs 
measured and the error ($\sigma$) to be the maximum deviation between the
two/three values from the mean, of multiplied 
by 0.63 \citep[small sample statistics; see][]{kee62}.

Membership assessment was performed by looking at the radial velocity distribution only, and assigning
individual star membership to a cluster when the star radial velocity is within 2$\sigma$ from the cluster
mean radial velocity. By adopting this criterion, stars are classified
as members (M) or not members (NM) in the last column
of Table 2. In most cases we found that the observed giants were cluster members.  

We compared the stars radial velocity with the literature, and found the
following:

\begin{description}

\item {\bf Collinder 110}: \cite{pan10} report
38.74$\pm$0.64 km s$^{-1}$
for star $\#$2129, which is very close to our estimate
(see Table~2). The lower resolution study by \cite{car07} suggest
a mean cluster velocity of 45$\pm$8 km s$^{-1}$ from 8 stars.
This value is again in fine agreement with \cite{pan10}
and this study.

\item {\bf NGC 2506} Star $\#$3231 was measured by \cite{red12}.
Their value (84.9$\pm$0.4km s$^{-1}$)  is in fine agreement with our.
Besides, except for star $\#$5271, all our program stars have
measurements in \cite{mer08}. Our values are in fine agreement
for all the common stars. In particular, stars $\#$2109, that we
considered a non-member, has a very different radial velocity in 
\cite{mer08}. Its velocity (80.92 km s$^{-1}$)
confirms it is most probably a binary stars.

\item{\bf Collinder 261}: We do not have any star in common with
\cite{car05}, however our radial velocities are fully compatible
with that study. $\#$ 2291 and 2311 are in common with
\cite{sil07}, and their values (--27.8 and --18.1)
are only in marginal agreement with our study.
\cite{sil07} are, however, based only on a narrow
spectral range, and are affected by errors as a large as 2 km s$^{-1}$.

\item{\bf  NGC 2477}: \cite{mer08} measured radial velocity
for  83 stars in NGC~2477.  They obtained 7.26$\pm$1.00 km s$^{-1}$ as
cluster mean radial velocity. Our program stars have compatible radial
velocity, and  support the non member nature of stars $\#$ 5043 and 8216.

\item {\bf NGC 5822}:  the most recent radial velocity study is from
\cite{mer08}.
These authors derive a mean radial velocity of --29.31$\pm$0.82 km s$^{-1}$
from 28 stars, and this is in nice agreement with our values.
This confirms our classification as non-member of stars $\#$16450.
 
\end{description}

In conclusion, the agreement with literature values is in general very good.

The processing of spectra (continuum definition,
equivalent widths measurements etc.) was carried out using the
DECH20 software package \citep{gal92}.
The results of the comparison of the equivalent widths of the lines measured
in this work with the ones measured by other authors for two giant stars
are the following: Cr~110 (star 2129, \cite{pan10}) ,
$<$EW(our) -- EW(lit)$>$ = 0.11 $\pm$4.44 m\AA\ (162 lines)
and  NGC~2506 (star 3231, \cite{red12}),
$<$EW(our)--EW(lit)$>$ = 0.05 $\pm$3.64 m\AA\ (116 lines).
This is illustrated in Figs.1 panels, from which one can appreciate
the  good agreement between the different measurement systems.

\begin{figure*}
\begin{tabular}{cc}
\includegraphics[width=8cm]{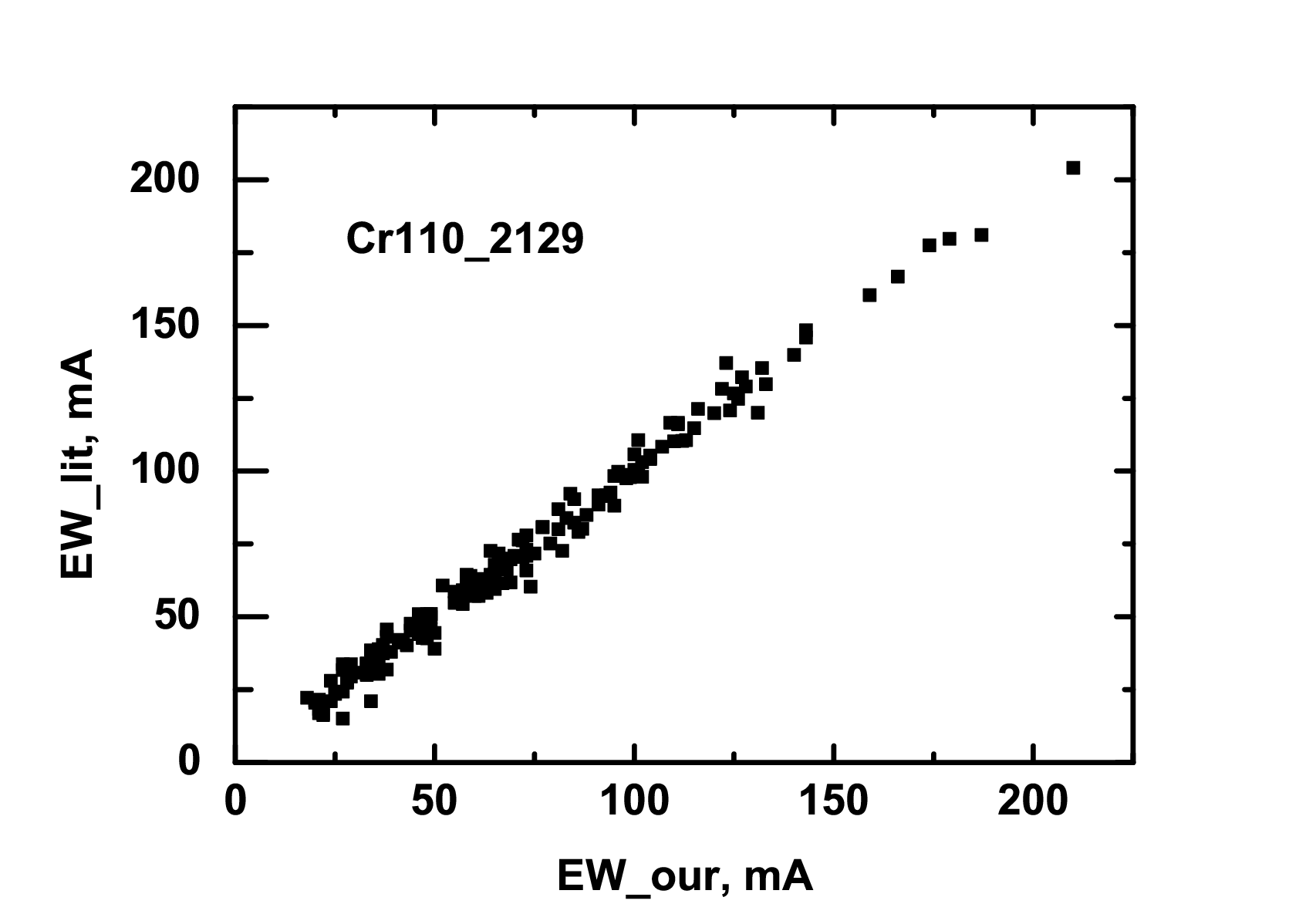}&\includegraphics[width=8cm]{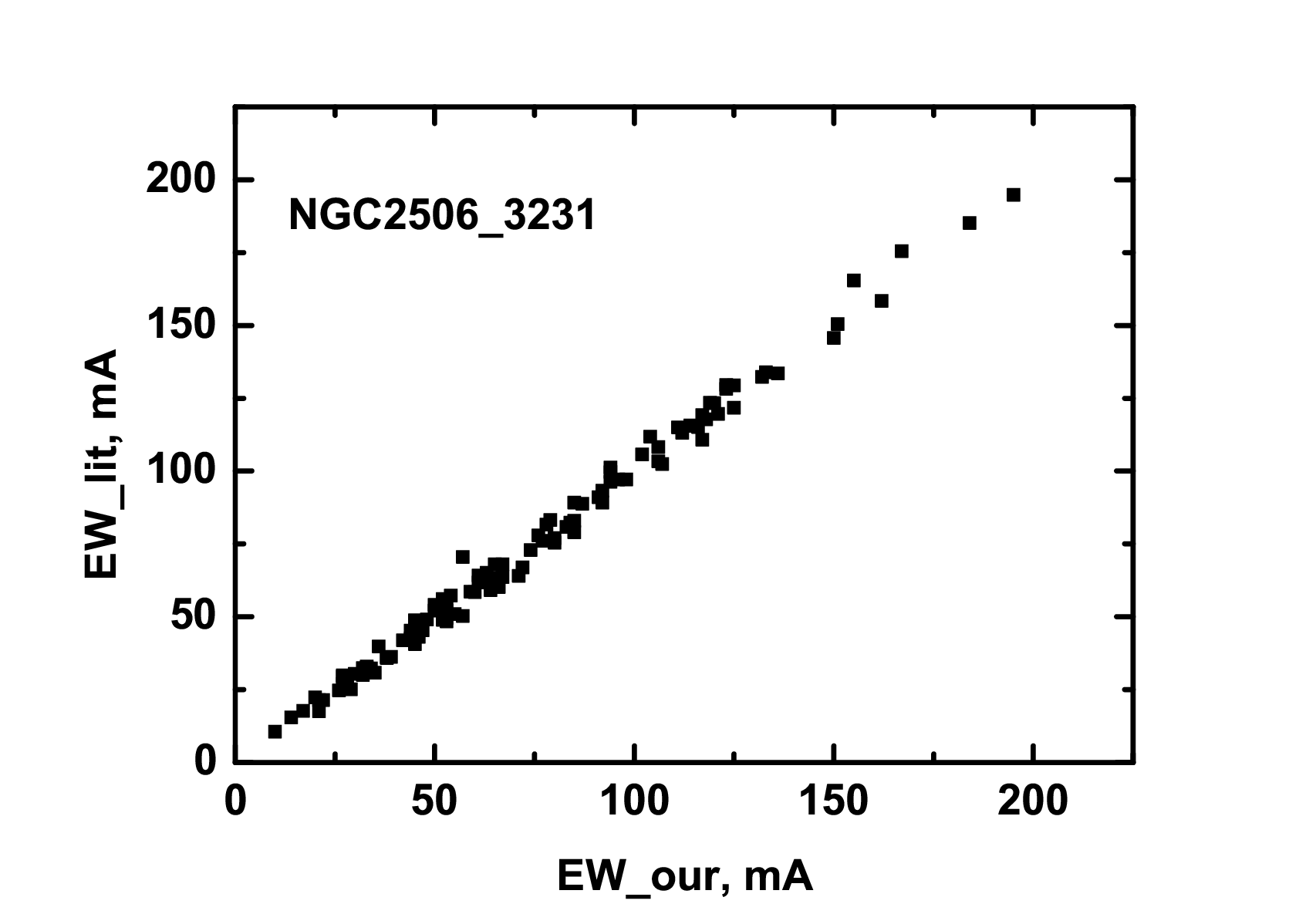}\\
\end{tabular}
\caption{The comparison of the equivalent widths for star
  Cr~110$\_$2129 and NGC~2506$\_$3231 with literature data.}
\label{ewr4}
\end{figure*}

\begin{figure*}
\begin{tabular}{ccc}
\includegraphics[width=6cm]{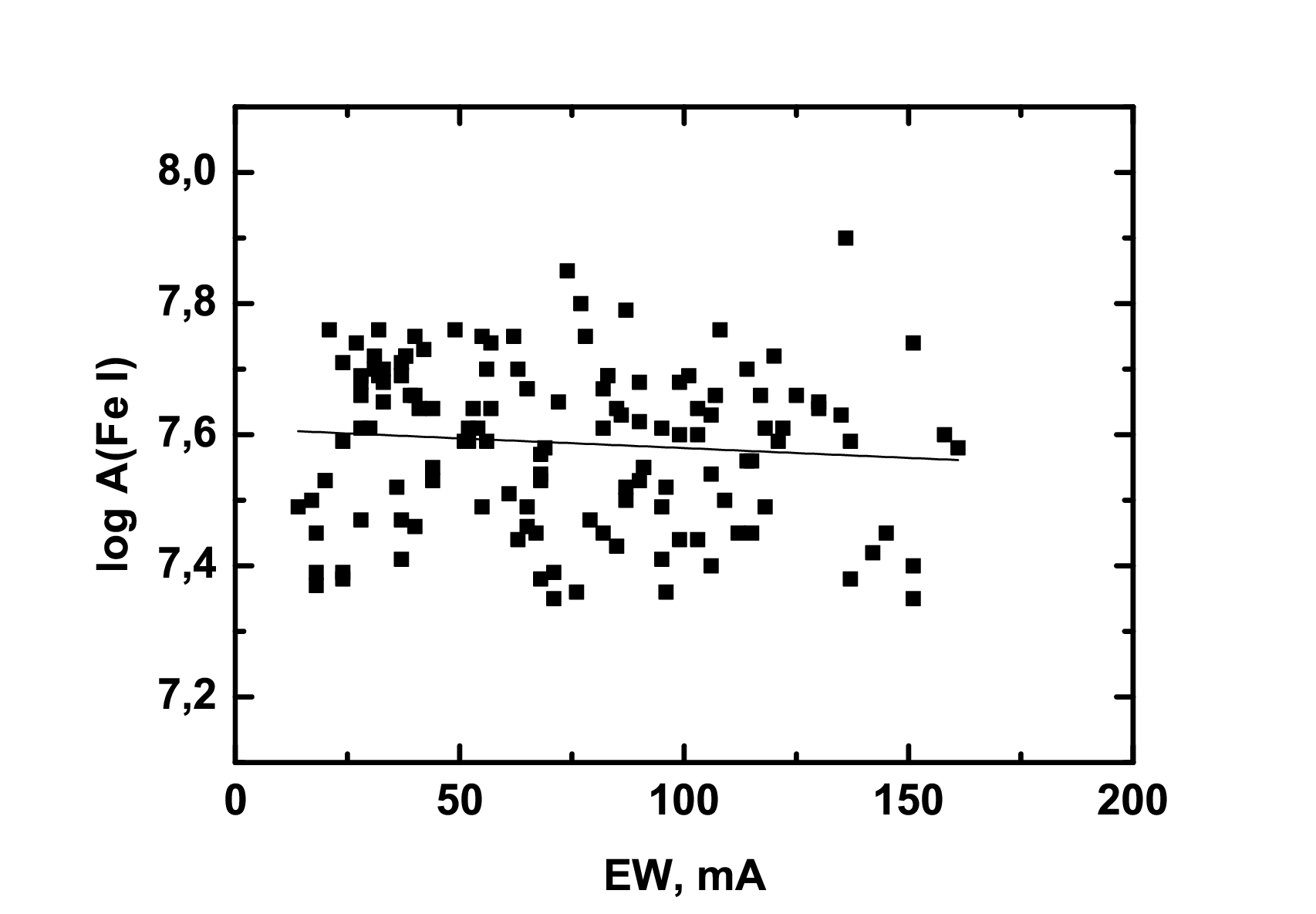}&\includegraphics[width=6cm]{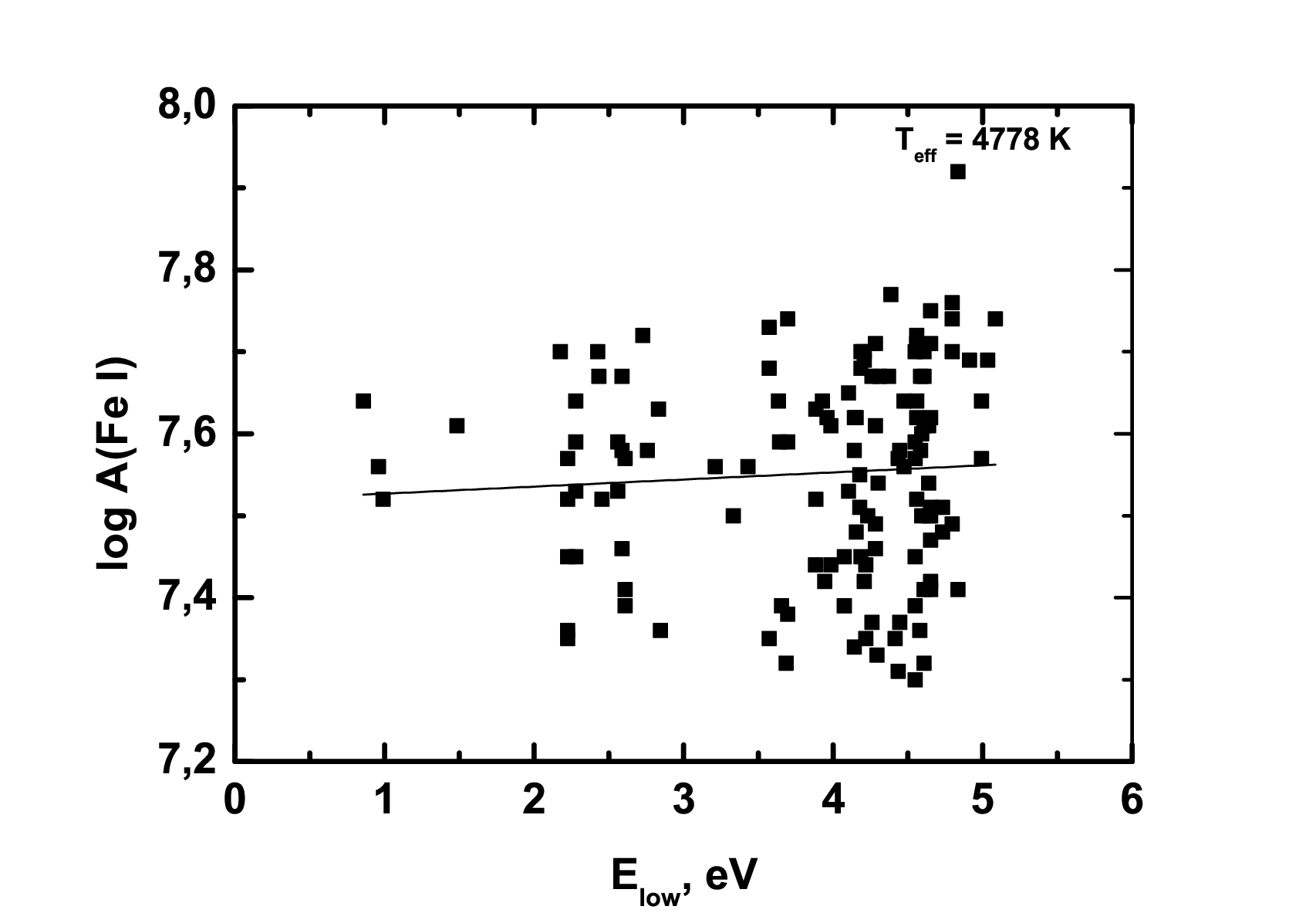}&\includegraphics[width=6cm]{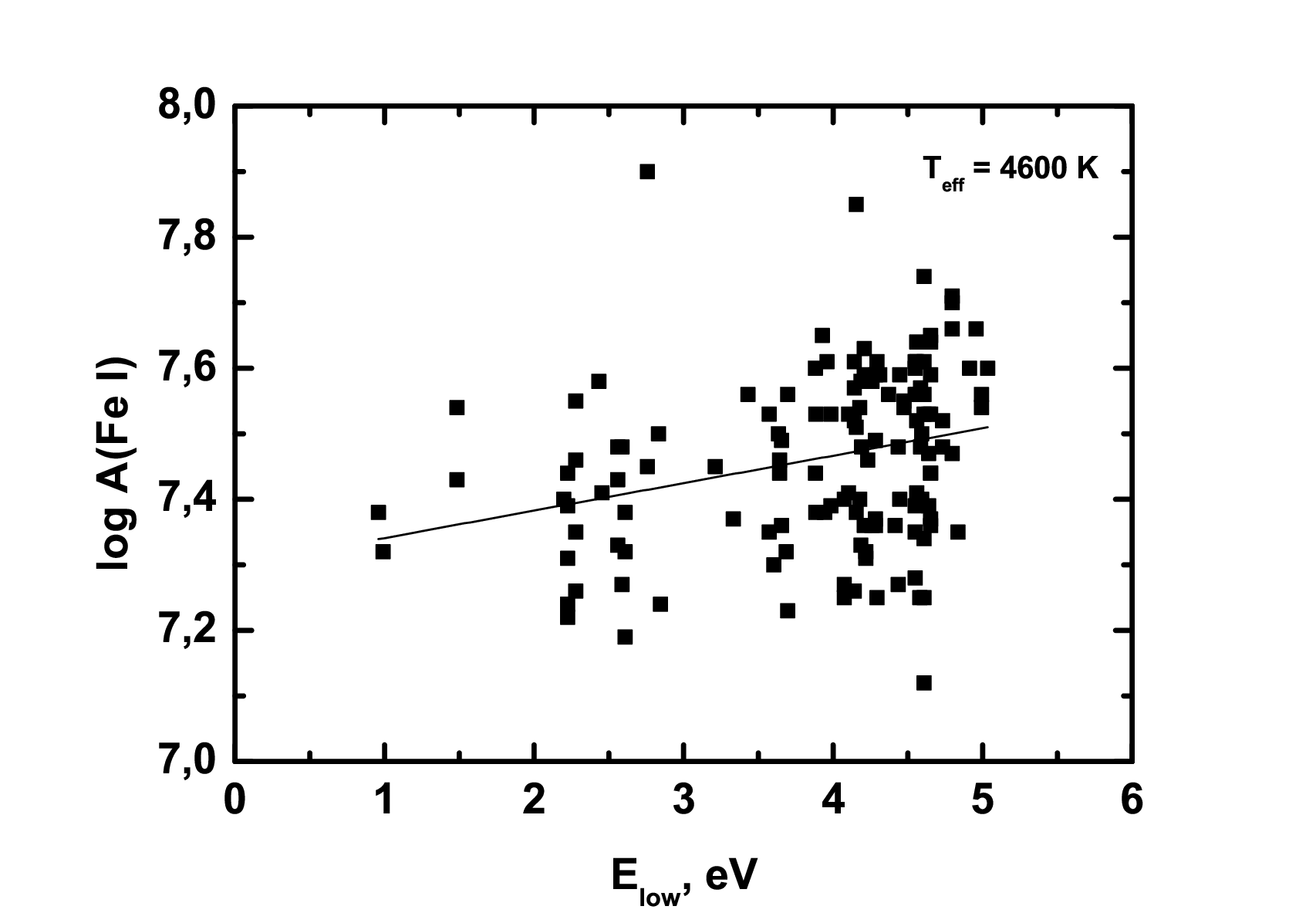}\\
\end{tabular}
\caption{For star  Cr~261$\_$2311,
the dependence of  the iron abundance
(based on Fe I lines) on the equivalent width
EW (choice of turbulent velocity \Vt, left) and
a similar dependence of the iron abundance on
the potential of the lower level of the line
$E_{low}$ for two values of the effective temperature \Teff\ (middle and right).}
\label{ewr7}
\end{figure*}

\section{Stellar atmospheric parameters}

Stars' effective temperatures \Teff\  were estimated by calibrating
the ratio of the central depths of the lines with different potentials of the
lower levels developed by \cite{ko06}.
The surface gravities \logg\ were computed using the iron ionization
balance.
The micro-turbulence velocity V${\rm t}$ was derived considering that the
iron abundance log A(Fe) obtained from the given  Fe~{\sc i} line is not
correlated with the equivalent width (EW) of that line.
The adopted value of the metallicity [Fe/H] is calculated using the iron
abundance obtained from  Fe~{\sc i} lines.
The resulting atmospheric parameters are presented in Table 2.

The comparison of the atmospheric parameters with literature data is
presented in Table 3.
One can notice that the external accuracy of the effective temperature \Teff\
is within
$\Delta$\Teff =  $\pm$100 K, the surface gravity
\logg -- $\Delta$\logg = $\pm$0.2 dex,
except the star Cr~261$\_$2311.
The difference in \Teff\ for this star reaches 178 K,  and 0.5 for gravity.
To check the choice of the temperatures we investigated
 dependences of  iron abundances log A(Fe~{\sc i})
determined using the Fe~{\sc i} lines on its excitation potential of low
level  and on EW for two micro-turbulence velocities
and  for two models with \Teff = 4748 K (our determination) and
\Teff = 4600 K \citep{sil07}.
This is shown in Fig~2 panels, from which one can appreciate
the lack of any clear trend.

The comparison with the C14 study for the stars observed in NGC~2477
and NGC~5822
is presented in the second part of Table 3. The agreement is generally good,
with a maximum difference in \Teff\ and \logg\ of 109\,K and 0.3\,dex,
for star \#2397 and \#18897 in NGC\,5822,
respectively. We derive an iron content on average $\sim$0.04\,dex and $\sim$0.12\,dex higher than C14, 
for stars in NGC\,5822 and NGC\,2477, respectively. The maximum differences are noted for 
NGC~2477 stars \#4221 (0.16\,dex),\#5076 (0.16\,dex), and \#7273 (0.23\,dex), C14 adopted a solar iron
abundance of 7.52, compared to the 7.57 adopted here. For the sake of an easier comparison, 
we reported absolute iron abundances in Table 3. Several authors have investigated in the literature 
differences in the derived abundances and chemical parameters as
estimated by different researchers adopting
different prescriptions and approaches \citep[see, e.g., ][]{bensby09,gilmore13}. We consider 
the agreement with C14 analysis as satisfactory. It provides as well an estimate of the 
differences which one expects from the analysis performed by different authors.
Note that the mentioned differences are not affecting the results of this work.

\begin{table*}
\caption{Comparison of atmospheric parameters.}
\label{compar}
\begin{tabular}{lcccccccc}
\hline
Star        &  \Teff ,K &  \logg &  \Vt &  [Fe/H]  &\Teff, K&\logg & \Vt & [Fe/H]  \\
                          &              &  lit      &       &                                &       & this study &       \\
\hline
Cr 110 2129 \citep{pan10} &   4950   &  2.7  &  1.4 &  0.05 & 4933 &  2.6 & 1.2  &--0.04\\
NGC 2506  \citep{red12}  &   5000   &  2.5  &  1.4 &--0.25 & 4974 &  2.6 & 1.2  &--0.22 \\
\hline
        &  \Teff ,K &  \logg &  \Vt &   log A(Fe) &\Teff, K&\logg & \Vt &  log A(Fe) \\
\hline
Cr 261 2291 \citep{sil07}&   4650   &  2.3  &  1.8 &  7.51 & 4746 &  2.5 & 1.2  &  7.57 \\
Cr 261 2311 \citep{sil07}&   4600   &  2.0  &  0.9 &  7.56 & 4778 &  2.5 & 1.15 &  7.55 \\
NGC\,2477 4027  (C14) & 4998 & 2.78 & 1.12 &  7.66 &  4966  &  2.7  &   1.4    &  7.67 \\ 
NGC\,2477 4221  (C14) & 4956 & 2.70 & 1.12 &  7.60 &  4975  &  2.8  &   1.2    &  7.76 \\ 
NGC\,2477 5043  (C14) & 5075 & 2.96 & 1.06 &  7.56 &  5001  &  2.8  &   1.2    &  7.65 \\ 
NGC\,2477 5076  (C14) & 5010 & 2.80 & 1.14 &  7.59 &  4954  &  2.7  &   1.2    &  7.75 \\ 
NGC\,2477 7266  (C14) & 5036 & 2.92 & 1.09 &  7.65 &  4966  &  2.8  &   1.2    &  7.76 \\ 
NGC\,2477 7273  (C14) & 4977 & 2.67 & 1.20 &  7.54 &  4985  &  2.8  &   1.2    &  7.77 \\ 
NGC\,2477 8216  (C14) & 5017 & 2.84 & 0.99 &  7.64 &  4945  &  2.7  &   1.2    &  7.71 \\ 
NGC\,5822 13292 (C14) & 5066 & 2.80 & 1.12 &  7.57 &  5010  &  2.8  &   1.2    &  7.61 \\
NGC\,5822 16450 (C14) & 5017 & 2.71 & 1.09 &  7.51 &  4972  &  2.6  &   1.2    &  7.55 \\ 
NGC\,5822 18897 (C14) & 5115 & 3.00 & 1.10 &  7.50 &  5030  &  2.7  &   1.0    &  7.55 \\ 
NGC\,5822 2397  (C14) & 5145 & 2.95 & 1.13 &  7.57 &  5036  &  2.8  &   1.1    &  7.59 \\ 
\hline
\end{tabular}
\end{table*}

\section{Abundance analysis}

The abundances of the investigated elements
are determined for 27 giants using the LTE approximation, and atmosphere
models by \cite{ck04}, computed for
the parameters of each star.
The estimate of the oxygen and Eu abundance was
performed with a new version of the STARSP software
package \citep{tsy96}.  For this we used the line list in the region
of the [{O}~{\sc I}] line 6300.3 \AA\  and the europium line 6645.13 \AA\
from the VALD atomic data \citep{kup99}.

The magnesium, sodium, and barium abundances were computed in  NLTE  approximation with a version
of MULTI \citep{car86}, modified by S. Korotin \citep[][]{mis04,kor99,kor11} .
We used the Mg~{\sc i} lines 5172.69, 5183.61, 5528.41, 5711.09, 6318.7, 6319,24, 6319.49~\AA\AA;
the NaI 5682.65, 5688.22, 6154.23, 6160.75  \AA\AA\ and
three lines of Ba~{\sc ii}(5853, 6141 and 6496 \AA\AA).

The model of sodium atom consists of 27 levels of Na~{\sc i} and the ground level
of Na~{\sc i}. We considered the radiative transitions between the
first 20 levels of Na~{\sc i} and the ground level of Na~{\sc ii}.
Transitions between the remaining levels were used only in the equations
of particle number conservation. Finally,
46 $b-b$ and 20 $b-f$ transitions were included in the linearisation procedure.
The NLTE corrections for the Na abundances are $\aplt$0.2 dex.

We employed the model of magnesium atom consisting of 97 levels: 84 levels of
Mg~{\sc i}, 12 levels of Mg~{\sc ii} and a ground state of Mg~{\sc iii}.
Within the described system of the magnesium atom levels, we considered the
radiative transitions between the first 59 levels of  Mg~{\sc i} and ground
level of Mg~{\sc ii}. Transitions between the rest levels were not taken
into account and they were used only in the equations of particle number
conservation.
The NLTE corrections for the Mg abundances are $\aplt$0.1 dex.
Our Ba model contains 31 levels
of Ba~{\sc i}, 101 levels of Ba~{\sc ii}  with $n < 50$, and the
ground level of Ba~{\sc iii} ion.
We also included 91 bound-bound transitions.
The odd Ba isotopes have hyperfine splitting of their levels and, thus,
several Hyper Fine Structure (HFS) components for each line \citep{rut78}.
Therefore, line 6496 \AA\ was fitted by adopting the
even-to-odd abundance ratio of 82:18 \citep{cam82}.
The HFS for lines 5853\AA\ and 6141\AA\ is not significant. The solar Ba
abundance was assumed to be $(Ba/H)_{\odot}$ =  2.17 where  log A(H) = 12.
That value was obtained from the Solar Atlas \citep{kur84} with the
same atomic data, which had been used to estimate the Ba abundance in the
stellar atmospheres. The influence of the NLTE does not have
any significant effect in the examined stars. The NLTE corrections for the
Ba abundances are $\aplt$0.1 dex.

\subsection{Errors in abundance determinations}

The effects of uncertainties in atmospheric parameters on the accuracy of
elemental abundance determinations
 for star NGC~2477$\_$7266 is given in Table \ref{errors}. The typical errors in
temperature \Teff,  surface gravity \logg\, and
microturbulent velocity \Vt\, are $\pm$100 K (col 1), $\pm$0.2 (col 2)
and $\pm$0.2 km s$^{-1}$ (col 3), respectively.
The total error (col 4) includes the mean error in the equivalent width
measurements and the accuracy of
the synthetic spectrum fitting that is assumed to be 0.05 dex.

\begin{table}
\caption{Abundance uncertainties due to atmospheric parameters.
 NGC~2477$\_$7266 (\Teff=4966, \logg=2.8, \Vt=1.2, [Fe/H] = 0.19).}
\label{errors}
\begin{tabular}{lcccc}
\hline
Species  &   $\Delta$ \Teff+100  & $\Delta$ \logg+0.2 & $\Delta$ \Vt+0.02 & Total \\
\hline
 O {\sc i}   &  0.02  &  0.08 &   0.00 &   0.09  \\
Na {\sc i}   &  0.08  &  0.01 &   0.02 &   0.10  \\
Mg {\sc i}   &  0.05  &  0.01 &   0.04 &   0.08  \\
Al {\sc i}   &  0.07  &  0.01 &   0.03 &   0.09  \\
Si {\sc i}   &  0.02  &  0.01 &   0.03 &   0.06  \\
Ca {\sc i}   &  0.09  &  0.03 &   0.09 &   0.14  \\
Sc {\sc ii}  &  0.01  &  0.07 &   0.05 &   0.10  \\
Ti {\sc i }  &  0.13  &  0.01 &   0.04 &   0.14  \\
Ti {\sc ii}  &  0.01  &  0.07 &   0.07 &   0.11  \\
V {\sc i  }  &  0.14  &  0.00 &   0.05 &   0.16  \\
Cr {\sc i }  &  0.09  &  0.01 &   0.05 &   0.11  \\
Fe {\sc i }  &  0.07  &  0.01 &   0.07 &   0.11  \\
Fe {\sc ii}  &  0.08  &  0.06 &   0.07 &   0.13  \\
Co {\sc i }  &  0.07  &  0.03 &   0.04 &   0.10  \\
Ni {\sc i }  &  0.05  &  0.02 &   0.05 &   0.09  \\
Y {\sc ii }  &  0.00  &  0.07 &   0.10 &   0.13  \\
Zr {\sc ii}  &  0.01  &  0.08 &   0.02 &   0.10  \\
Ba {\sc ii}  &  0.02  &  0.03 &   0.14 &   0.16  \\
La {\sc ii}  &  0.02  &  0.08 &   0.05 &   0.11  \\
Ce {\sc ii}  &  0.01  &  0.08 &   0.01 &   0.10  \\
Nd {\sc ii}  &  0.01  &  0.08 &   0.04 &   0.10  \\
Eu {\sc ii}  &  0.01  &  0.07 &   0.04 &   0.10  \\
 \hline
\end{tabular}
\end{table}

As can be seen from Table \ref{errors}, the total error in the elemental
abundance determinations is less than 0.2 dex. In particular, the error
associated to the determination of the Ba abundances is 0.16 dex.

The solar abundance computed for the lines from the Solar spectrum \citep{kur84}
with log gf from VALD data base \citep{kup99} and the solar model \citep{ck04} 
 is given in Table~\ref{solar}.
The elemental abundances obtained by us for studied OCs,
namely Cr~110, Cr~261, NGC~2477,
NGC~2506 and NGC~5822 are given in Tables 6--10 and
the mean abundance values for each cluster are presented in Table 11.

Since the Ba overabundance found for a number of OCs is the most
controversial result from recent spectroscopic observations of OCs,
here below we discuss possible source of uncertainties that may affect
Ba measurements. In particular, let us consider possible causes of the
Ba overabundance resulted from the equivalent width (EW's) measurements,
applying methods of abundance determination, such as growth curve or
synthetic spectrum techniques under
both the LTE and non-LTE approximations, usage of different atmospheric
model grids, etc.

The investigated Ba~II lines (4554, 5853, 6141 and 6496 \AA\AA) tend to
be strong (ranging from 100 to 450  m\AA) in the spectra of OC giants.
In this case it is crucial to correctly account for the wings of spectral lines, i.e. to establish the continuous spectrum level.
That may cause errors in the EW measurements of up to 10--15\%.
It is especially important when measuring the equivalent widths
of lines or when applying the growth curve technique.
Moreover, improper consideration of spectral line damping constants, especially the van der Waals broadening, can result
in additional error. However as the Ba lines are wide enough, their profiles are affected by blending of other lines.
The 6141 \AA\ line blending affects the central part of the line (Fe I line). The effects of such distortions (blending) in spectral line
profiles can be taken into account only when calculating the synthetic spectrum.

The estimates obtained in the study by \cite{mis13b}
(Figs. 5, 6 from that paper) indicate that the equivalent widths
and profiles are rather sensitive to the Ba abundance.
Relatively weaker and moderate lines (up to 200 m\AA~) are very
sensitive to the elemental abundance changes; whereas stronger lines
when using the computed synthetic spectra, allow to obtain
the abundance values with an accuracy of not less than $\pm$0.1 dex
(see Figs. 6, 7). Applying different atmospheric model grids
can also cause some uncertainty in the abundance determinations of
up to 0.05--0.1 dex.

In the paper \cite{dor12} the authors suggest several possible explanations
for the Ba overabundance, such as:
(1) neglecting the hyperfine structure of the Ba lines;
(2) deviations from the Local Thermodynamic Equilibrium (LTE) conditions;
(3) the chromospheric activity (see also \cite{dor09}).

In the  OCs studied by \cite{mis13b} and in this work,
the Ba lines are
strong and broad. Therefore, neglecting of the hyperfine structure is
not relevant in this case. Concerning the second point mentioned by
\cite{dor12}, we considered NLTE correction for the Ba analysis.
Note also that 
LTE deviations that we found do not exceed 0.1 dex. Finally, concerning the
chromospheric activity, already \cite{dor12}  did not find any correlation
between the Ba abundance and the chromospheric activity indices
for the investigated stars. We would therefore consider the impact of this
last source of uncertainty as marginal.

Thus, the definition of  the continuous spectrum level, the EW measurement
errors, the usage different abundance determination
techniques, the adoption of  atomic parameters (damping constants) and various
atmospheric models can result in uncertainties in
Ba abundance estimates obtained by different authors of up
to $\sim$0.2 dex.

However, we did find variations up to 0.3 dex between
different references in the literature, and in one case almost 0.4 dex
(see Section~5).

\begin{table}
\caption{Solar abundance derived by us and compared with photospheric
    abundance by Asplund et al. (2009).}
\label{solar}
\begin{tabular}{llrl}
\hline
Species &    log A (this work)  & NL   &\cite{as09}   \\
 \hline
 O {\sc i}    &  {\bf 8.70}        &  1    & 8.69 $\pm$0.05  \\
Na {\sc i}    &{\bf 6.25 $\pm$0.04}&  10    &  6.24 $\pm$0.04  \\
Mg {\sc i}    &{\bf 7.58 $\pm$0.02}&  9    &  7.60 $\pm$0.04  \\
Al {\sc i}    &    6.30 $\pm$0.01 &  2     &  6.45 $\pm$0.03  \\
Si {\sc i}    &    7.55 $\pm$0.08 &  23    &  7.51 $\pm$0.03  \\
Ca {\sc i}    &    6.32 $\pm$0.07 &  16    &  6.34 $\pm$0.04  \\
Sc {\sc i}    &          ...      &       &  3.15 $\pm$0.04  \\
Sc {\sc ii}   &    3.22 $\pm$0.11 & 14    &        ...       \\
Ti {\sc i }   &    4.96 $\pm$0.08 & 41    &  4.95 $\pm$0.05  \\
Ti {\sc ii}   &    5.01 $\pm$0.03 & 5     &        ...       \\
V {\sc i  }   &    4.04 $\pm$0.12 & 36    &  3.93 $\pm$0.08  \\
Cr {\sc i }   &    5.67 $\pm$0.09 & 23    &  5.64 $\pm$0.04  \\
Fe {\sc i }   &    7.57 $\pm$0.08 & 164   &  7.50 $\pm$0.04  \\
Fe {\sc ii}   &    7.47 $\pm$0.04 & 11    &        ...       \\
Co {\sc i }   &    5.00 $\pm$0.10 & 28    &  4.99 $\pm$0.07  \\
Ni {\sc i }   &    6.29 $\pm$0.06 &  56   &  6.22 $\pm$0.04  \\
Y {\sc ii }   &    2.15 $\pm$0.17 &  7    &  2.21 $\pm$0.05  \\
Zr {\sc i }   &          ...      &       &  2.58 $\pm$0.04  \\
Zr {\sc ii}   &    2.79  $\pm$0.19&  2    &        ...       \\
Ba {\sc ii}   &{\bf 2.17$\pm$0.04}&  4    &  2.18$\pm$0.09   \\
La {\sc ii}   &    1.24 $\pm$0.02 &  2     &  1.10 $\pm$0.04  \\
Ce {\sc ii}   &    1.70 $\pm$0.11 &  6     &  1.58 $\pm$0.04  \\
Nd {\sc ii}   &    1.54 $\pm$0.08 &  11   &  1.42 $\pm$0.04  \\
Eu {\sc ii}   &    {\bf 0.60}     &  1       &  0.52 $\pm$0.04  \\
\hline
\end{tabular}
  \\
Notes. The abundance values computed with synthetic spectrum marked as bold.
\end{table}

\begin{table*}
\caption{Abundance results for Cr 110.}
\label{cr110}
\tiny
\begin{tabular}{lrrrrrrrrrrrrrrrrrr}
\hline
    & & 1122    & & &                 1134   & & &                   1149     & & &             1151 & & &               2129  & & &                 3132& \\
\hline
  Ion& [El/H]& $\sigma$ & NL &   [El/H] & $\sigma$ & NL &  [El/H] & $\sigma$ & NL &  [El/H] & $\sigma$ & NL  &[El/H] & $\sigma$ & NL &[El/H] & $\sigma$ & NL\\
\hline
 O {\sc i}  &   {\bf0.05}& ...&   1&       {\bf0.15  }&...&   1&         {\bf0.05}&...&    1&         {\bf0.05}&...&   1&       {\bf0.05}&...&   1&    {\bf--0.05}&...&     1  \\
Na {\sc i}  &  {\bf--0.05}&...&   4&       {\bf--0.03}&...&   4&         {\bf0.00}&...&   4&         {\bf0.00}&...&   4&     {\bf--0.02}&...&  4&    {\bf--0.03}&...&     4  \\
Mg {\sc i}  &  {\bf--0.06}&...&   7&       {\bf--0.06  }&...& 7&         {\bf--0.08}&...& 7&         {\bf--0.04}&...& 7&    {\bf--0.07}&...&   7&    {\bf--0.11}&...&     7  \\
Al {\sc i}  &  --0.06&  0.01&     2&        0.05&  0.06&     2&        --0.02&  0.05&     2&         0.06&  0.04&     2&       0.10&  0.05&     2&       0.09&  0.02&     2  \\
Si {\sc i}  &  --0.02&  0.09&    14&        0.04&  0.11&    17&          0.06&  0.16&    21&         0.08&  0.13&    18&       0.00&  0.13&    17&       0.00&  0.11&    20  \\
Ca {\sc i}  &  --0.04&  0.08&    11&        0.04&  0.14&    11&        --0.05&  0.14&    13&         0.07&  0.12&    12&     --0.06&  0.11&    12&     --0.07&  0.08&    14  \\
Sc {\sc ii} &  --0.07&  0.17&     9&        0.08&  0.14&     7&          0.00&  0.18&    10&         0.01&  0.14&     9&     --0.09&  0.07&     8&     --0.07&  0.13&    10  \\
Ti {\sc i } &  --0.05&  0.09&    26&      --0.04&  0.10&    31&        --0.06&  0.08&    34&       --0.01&  0.10&    34&     --0.04&  0.08&    25&     --0.04&  0.12&    30  \\
Ti {\sc ii} &  --0.06&  0.10&     4&      --0.05&  0.08&     3&          0.02&  0.13&     5&         0.06&  0.10&     4&     --0.09&  0.14&     4&       0.08&  0.11&     4  \\
V {\sc i  } &  --0.07&  0.08&    15&      --0.00&  0.09&    15&        --0.07&  0.08&    20&       --0.05&  0.10&    28&     --0.03&  0.09&    19&     --0.05&  0.13&    28  \\
Cr {\sc i } &  --0.06&  0.07&    12&        0.00&  0.14&    18&        --0.09&  0.09&     8&         0.07&  0.10&    15&     --0.01&  0.06&     7&     --0.05&  0.04&    10  \\
Fe {\sc i } &  --0.06&  0.11&   112&        0.02&  0.12&   120&        --0.01&  0.12&   136&         0.02&  0.11&   130&     --0.04&  0.09&   119&     --0.03&  0.12&   134  \\
Fe {\sc ii} &  --0.11&  0.05&     7&      --0.07&  0.07&     8&        --0.10&  0.14&    11&       --0.11&  0.11&     8&     --0.07&  0.07&     7&     --0.13&  0.10&     8  \\
Co {\sc i } &  --0.02&  0.04&    12&      --0.03&  0.10&    23&        --0.10&  0.07&    18&       --0.05&  0.12&    21&     --0.05&  0.10&    13&     --0.07&  0.10&    16  \\
Ni {\sc i } &  --0.06&  0.08&    30&      --0.04&  0.09&    39&        --0.08&  0.10&    45&       --0.02&  0.11&    45&     --0.09&  0.07&    37&     --0.07&  0.13&    47  \\
Y {\sc ii } &  --0.04&  0.15&     8&      --0.00&  0.14&     7&          0.11&  0.12&     8&         0.17&  0.04&     7&       0.01&  0.14&     8&       0.10&  0.14&     6  \\
Zr {\sc ii} &  --0.06&  0.01&     3&        0.14&  0.25&     3&          0.10&  0.15&     3&         0.11&  0.03&     3&     --0.09&  0.18&     3&       0.00&  0.15&     3  \\
Ba {\sc ii} &  {\bf0.27}& ...&   3&       {\bf0.33}&...&     3&         {\bf0.38}&...&    3&         {\bf0.31}&...&   3&       {\bf0.31}&...&   31&      {\bf0.36}&...&   3  \\
La {\sc ii} &    0.09&  0.08&     2&        0.07&  0.02&     2&          0.19&  0.06&     2&         0.19&  0.13&     3&       0.10&  0.05&     3&       0.24&  0.10&     2  \\
Ce {\sc ii} &    0.07&  0.16&     6&        0.21&  0.16&     5&          0.22&  0.14&     7&         0.06&  0.16&     4&       0.09&  0.09&     3&       0.02&  0.12&     6  \\
Nd {\sc ii} &    0.09&  0.14&    10&        0.16&  0.11&     6&          0.14&  0.10&     9&         0.08&  0.11&     8&       0.03&  0.11&     9&       0.02&  0.07&     9  \\
Eu {\sc ii} & {\bf0.25}&... &     1&      {\bf0.20}&...&     1&         {\bf0.20}&...&    1&        {\bf0.23}&...&    1&      {\bf0.16}&...&    1&      {\bf0.16}&...&    1  \\
\hline
\end{tabular}
  \\
Notes. The abundance values computed with synthetic spectrum marked as bold.
 \end{table*}

\begin{table*}
\caption{Abundance results of Cr 261.}
\label{cr261}
\tiny
\begin{tabular}{lrrrrrrrrrrrrrrr}
\hline
    & & 2269    & & &                     2291   & & &                 2309     & & &                2311 & & &                   2313  & \\
\hline
  Ion& [El/H]& $\sigma$ & NL &   [El/H] & $\sigma$ & NL &  [El/H] & $\sigma$ & NL &  [El/H] & $\sigma$ & NL  &[El/H] & $\sigma$ & NL \\
\hline
Na {\sc i}   &    {\bf0.12}& ...&  4&       {\bf0.19  }&...&   4&         {\bf0.06}&...&    4&         {\bf0.14}&...&    4&       {\bf0.14}&...&   4   \\
Mg {\sc i}   &    {\bf0.07}& ...&  7&       {\bf0.03  }&...&   7&         {\bf--0.02}&...&   7&         {\bf0.04}&...&   7&       {\bf0.10}&...&   7   \\
Al {\sc i}   &    0.07&  0.10&     2&         0.15&  0.01&     2&        --0.01&  0.13&     2&          0.10&  0.11&     2&       0.18&  0.11&     2   \\
Si {\sc i}   &    0.07&  0.13&    15&         0.05&  0.12&    19&          0.08&  0.13&    16&          0.05&  0.11&    17&       0.03&  0.14&    18   \\
Ca {\sc i}   &    0.04&  0.08&    14&       --0.09&  0.11&     9&        --0.03&  0.12&    14&          0.00&  0.13&    13&     --0.09&  0.12&    11   \\
Sc {\sc ii}  &    0.06&  0.14&    11&         0.10&  0.14&     9&          0.10&  0.06&     6&          0.06&  0.11&    13&       0.10&  0.12&     8   \\
Ti {\sc i }  &    0.06&  0.13&    44&         0.05&  0.14&    41&          0.03&  0.15&    40&          0.03&  0.10&    40&       0.01&  0.13&    39   \\
Ti {\sc ii}  &    0.07&  0.06&     4&         0.09&  0.06&     4&          0.08&  0.07&     4&          0.01&  0.09&     4&       0.09&  0.06&     3   \\
V {\sc i  }  &    0.14&  0.10&    21&         0.10&  0.13&    27&          0.06&  0.12&    30&          0.05&  0.12&    28&       0.14&  0.14&    26   \\
Cr {\sc i }  &    0.06&  0.08&    19&       --0.00&  0.11&    14&        --0.01&  0.12&    15&          0.06&  0.13&    14&     --0.06&  0.08&    14   \\
Fe {\sc i }  &  --0.02&  0.11&   116&         0.00&  0.11&   133&          0.00&  0.12&   135&        --0.02&  0.12&   135&     --0.01&  0.11&   106   \\
Fe {\sc ii}  &  --0.08&  0.15&     7&       --0.04&  0.08&    12&        --0.07&  0.09&     8&        --0.09&  0.10&    10&     --0.08&  0.07&     9   \\
Co {\sc i }  &    0.03&  0.12&    16&         0.01&  0.16&    17&          0.05&  0.17&    19&          0.02&  0.14&    20&       0.05&  0.18&    21   \\
Ni {\sc i }  &    0.03&  0.11&    47&         0.02&  0.08&    45&          0.03&  0.13&    51&          0.08&  0.11&    52&       0.04&  0.13&    50   \\
Y {\sc ii }  &    0.09&  0.18&     6&         0.04&  0.18&     6&          0.11&  0.15&     7&          0.05&  0.19&     7&       0.06&  0.18&     3   \\
Zr {\sc ii}  &    0.01&  0.16&     3&         0.03&  0.10&     3&          0.13&  0.04&     2&          0.08&  0.10&     2&       0.11&  0.22&     2   \\
Ba {\sc ii}  &    {\bf0.14}& ...&  3&       {\bf0.40}&...&     3&         {\bf0.33}&...&     3&         {\bf0.36}&...&   3&       {\bf0.36}&...&   3   \\
La {\sc ii}  &    0.12&  0.01&     2&         0.11&  0.07&     4&          0.18&  0.02&     2&          0.15&  0.09&     2&       0.18&  0.11&     2   \\
Ce {\sc ii}  &    0.01&  0.08&     4&         0.05&  0.19&     7&         -0.00&  0.16&     6&          0.05&  0.17&     8&       0.14&  0.18&     8   \\
Nd {\sc ii}  &    0.05&  0.16&     9&         0.07&  0.16&     9&          0.02&  0.11&    10&          0.10&  0.12&     8&       0.08&  0.16&    11   \\
Eu {\sc ii}  &   {\bf0.20 }&...&   1&       {\bf0.27}&...&     1&         {\bf0.27}&...&    1&         {\bf0.20}&...&    1&       {\bf0.33?}&...&  1   \\
\hline
\end{tabular}
  \\
Notes. The abundance values computed with synthetic spectrum marked as bold.
 \end{table*}

\begin{table*}
\caption{ Abundance results for NGC 2477.}
\label{ngc2477}
\tiny
\begin{tabular}{lrrrrrrrrrrrrrr}
\hline
& \multicolumn{2}{c}{4027} & \multicolumn{2}{c}{4221}&\multicolumn{2}{c}{5043}&\multicolumn{2}{c}{5076} & \multicolumn{2}{c}{7266}&\multicolumn{2}{c}{7273}&\multicolumn{2}{c}{8216}\\
\hline
  Ion& [El/H]& $\sigma$,  NL &   [El/H] & $\sigma$,  NL &  [El/H] & $\sigma$,  NL &  [El/H] & $\sigma$,  NL & [El/H] & $\sigma$,  NL &  [El/H] & $\sigma$,  NL &  [El/H] & $\sigma$,  NL \\
\hline
 O {\sc i}   &{\bf--0.05}&...(1)& {\bf--0.15}& ...(1)&       {\bf--0.15}&...(1)&        {\bf--0.15}&...(1)&      {\bf--0.15}& ...(1)&     {\bf--0.15}&...(1)&   {\bf--0.15}&...(1) \\
Na {\sc i}   & {\bf0.13}&...(4)& {\bf0.13}& ...(4)&          {\bf0.06}& ...(4)&          {\bf0.10}&...(4)&         {\bf0.12}& ...(4)&       {\bf0.14}&...(4)&     {\bf0.09}&...(4)  \\
Mg {\sc i}   & {\bf--0.06}&...(7)7& {\bf--0.02}& ...(7) &       {\bf--0.09 }& ...(7)&     {\bf--0.01}&...(7)&      {\bf--0.01}& ...(7)&     {\bf--0.06}&...(7)&   {\bf--0.10}&...(7)  \\
Al {\sc i}   &  0.00&   0.12 (2)&    --0.12&  0.13(2)&      --0.19& 0.07(2)&     --0.07&  0.01(2)&     --0.11&  0.08(2)&     -0.09&   0.08(2)&    --0.10&   0.01(2) \\
Si {\sc i}   &  0.07&   0.17(21)&     0.18&   0.15(23)&     0.09&   0.16(23)&     0.17&   0.15(24)&     0.20&   0.17(23)&     0.20&   0.17(23)&     0.16&   0.20(21)  \\
Ca {\sc i}   &  0.00&   0.10(11)&     0.08&   0.19(16)&     0.00&   0.10(16)&     0.06&   0.17(15)&     0.09&   0.11(16)&     0.09&   0.10(16)&     0.07&   0.11(14) \\
Ti {\sc i }  &--0.04&   0.06(22)&     0.02&   0.09(3)&    --0.09&  0.07(26)&   --0.03&   0.08(38)&     0.02&   0.09(30)&     0.02&   0.10(34)&   --0.07&   0.09(33) \\
Ti {\sc ii}  &  0.03&   0.11 (4)&    --0.03&  0.09(3)&      0.07&   0.14(5)&      0.19&   0.18(5)&      0.13&   0.18(5)&      0.16&   0.17(5)&      0.14&   0.21(5)  \\
V {\sc i  }  &  0.01&   0.12(30)&     0.11&   0.14(34)&    --0.06&  0.10(32)&     0.08&   0.13(31)&     0.05&   0.13(32)&     0.08&   0.13(32)&   --0.01&   0.14(33) \\
Cr {\sc i }  &--0.01&   0.19(17)&     0.07&   0.14(19)&     0.00&   0.13(17)&     0.01&   0.09(16)&     0.05&   0.08(16)&     0.07&   0.08(16)&     0.05&   0.12(17)  \\
Fe {\sc i }  &  0.10&   0.12(127)&    0.19&   0.09(122)&    0.08&   0.12(134)&    0.18&   0.12(145)&    0.19&   0.12(146)&    0.20&   0.10(130)&    0.14&   0.12(138) \\
Fe {\sc ii}  &  0.05&   0.17 (8)&      0.16&  0.10(9)&      0.07&   0.05(10)&     0.11&   0.10(11)&      0.17&   0.18(9)&      0.18&   0.18(9)&     0.12&   0.06(10) \\
Co {\sc i }  &  0.11&   0.11(13)&     0.15&   0.13(18)&     0.03&   0.07(18)&     0.14&   0.12(19)&     0.15&   0.11(19)&     0.16&   0.10(17)&     0.08&   0.11(18) \\
Ni {\sc i }  &  0.06&   0.11(45)&     0.15&   0.10(49)&     0.02&   0.10(54)&     0.12&   0.10(47)&     0.16&   0.09(43)&     0.17&   0.11(5)&     0.08&   0.06(44)  \\
Y {\sc ii }  &  0.17&   0.09 (3)&      0.06&  0.12(4)&      0.10&   0.15(5)&      0.15&   0.17(5)&      0.07&   0.18(5)&      0.10&   0.18(5)&      0.15&   0.13(5)  \\
Zr {\sc ii}  &  0.03&   ... (1)&       0.10&  0.19(2)&       0.20&  ...(1)&        0.20&   0.18(2)&      0.18&  ...(1)&      0.22&  ...(1)&      0.27&   0.18(2)   \\
Ba {\sc ii}  & {\bf0.17}&... (3)& {\bf0.39}&    ...(3)&           {\bf0.39 }& ...(3)&  {\bf0.30}&...(3)&         {\bf0.30}& ...(3)&       {\bf0.26}&...(3)&     {\bf0.29}&...(3) \\
La {\sc ii}  &  0.09&  ... (1)&        0.30&   0.08(2)&      0.20&   0.14(2)&        0.22& ...(1)&      0.28&  ...     ( 1)&      0.20&   ...(1)&      0.19&   0.20(2)    \\
Ce {\sc ii}  &  0.11&   0.14 (4)&      0.26&   0.17(4)&      0.26&   0.14(6)&      0.22&   0.13(5)&      0.10&   0.18   ( 5)&      0.14&   0.18(5)&      0.22&   0.09(5)     \\
Nd {\sc ii}  &  0.06&   0.11 (3)&      0.00&   0.01(2)&      0.07&   0.16(9)&      0.04&   0.16(8)&      0.01&   0.17   ( 9)&      0.18&   0.18(9)&      0.00&   0.20(2)     \\
Eu {\sc ii}  & {\bf0.20}  &... (1)&{\bf0.15}&   ...(1)&   {\bf0.15}& ...(1)&     {\bf0.15}&...(1)&      {\bf0.18}& ...(1)&     {\bf0.15}&...(1)&   {\bf0.25}&...(1)  \\
\hline
\end{tabular}
  \\
Notes. The abundance values computed with synthetic spectrum marked as bold.
\end{table*}

\begin{table*}
\caption{Abundance results for NGC 2506.}
\label{ngc2506}
\tiny
\begin{tabular}{lrrrrrrrrrrrrrrrrrr}
\hline
& \multicolumn{3}{c}{1112} & \multicolumn{3}{c}{1229}&\multicolumn{3}{c}{2109}  & \multicolumn{3}{c}{2380} & \multicolumn{3}{c}{3231}&\multicolumn{3}{c}{5271} \\
\hline
  Ion& [El/H]& $\sigma$ & NL &   [El/H] & $\sigma$ & NL &  [El/H] & $\sigma$ & NL &  [El/H] & $\sigma$ & NL  &[El/H] & $\sigma$ & NL &[El/H] & $\sigma$ & NL\\
\hline
 O {\sc i}   &  {\bf--0.05}&--0&   1&      {\bf--0.10}& --&   1&     {\bf--0.10}& --&    1&       {\bf--0.05}& --&  1&     {\bf--0.10}& --&   1&       {\bf--0.00}& --&    1     \\
Na {\sc i}   &  {\bf--0.09}&--0&   4&      {\bf--0.21}& --&   4&     {\bf--0.21}& --&    4&       {\bf--0.13}& --&  4&     {\bf--0.16}& --&   4&       {\bf--0.13}& --&    4     \\
Mg {\sc i}   &  {\bf--0.19}&--0&   7&      {\bf--0.25}& --&   7&     {\bf--0.25}& --&    7&       {\bf--0.24}& --&  7&     {\bf--0.23}& --&   7&       {\bf--0.20}& --&    7     \\
Al {\sc i}   &    0.04&  0.01&     2&      --0.11&  0.07&     2&      --0.02&  0.11&     2&      --0.03&  0.02&     2&     --0.03&  0.03&     2&         0.01&  0.10&     2     \\
Si {\sc i}   &  --0.18&  0.09&    13&      --0.15&  0.07&    15&      --0.05&  0.16&    17&      --0.22&  0.14&    17&     --0.19&  0.13&    19&       --0.15&  0.13&    20     \\
Ca {\sc i}   &  --0.17&  0.10&    12&      --0.17&  0.08&    12&      --0.10&  0.19&     7&      --0.22&  0.11&    15&     --0.13&  0.08&    12&       --0.16&  0.08&    14     \\
Sc {\sc ii}  &  --0.16&  0.15&     8&      --0.13&  0.13&    13&      --0.18&  0.12&     9&      --0.07&  0.10&     8&     --0.07&  0.06&    11&       --0.08&  0.21&    13     \\
Ti {\sc i }  &  --0.28&  0.15&    37&      --0.28&  0.11&    38&      --0.19&  0.16&    24&      --0.17&  0.10&    27&     --0.24&  0.12&    33&       --0.27&  0.13&    31     \\
Ti {\sc ii}  &  --0.11&  0.03&     3&      --0.05&  0.08&     4&      --0.18&  0.12&     4&      --0.11&  0.01&     2&       0.02&  0.04&     2&       --0.01&  0.11&     4     \\
V {\sc i  }  &  --0.26&  0.07&    20&      --0.25&  0.11&    24&      --0.23&  0.09&    16&      --0.22&  0.10&    16&     --0.23&  0.14&    19&       --0.26&  0.07&    16     \\
Cr {\sc i }  &  --0.27&  0.15&    18&      --0.30&  0.08&    10&      --0.26&  0.19&    13&      --0.27&  0.13&     9&     --0.31&  0.13&    11&       --0.24&  0.12&     9     \\
Fe {\sc i }  &  --0.22&  0.10&   132&      --0.22&  0.11&   157&      --0.21&  0.17&    99&      --0.19&  0.13&   137&     --0.22&  0.13&   121&       --0.24&  0.12&   193     \\
Fe {\sc ii}  &  --0.28&  0.09&     5&      --0.27&  0.10&     9&      --0.26&  0.18&     4&      --0.25&  0.08&     6&     --0.28&  0.11&     9&       --0.28&  0.16&    13     \\
Co {\sc i }  &  --0.24&  0.14&    17&      --0.26&  0.13&    21&      --0.28&  0.17&    10&      --0.24&  0.14&    17&     --0.27&  0.14&    18&       --0.29&  0.14&    26     \\
Ni {\sc i }  &  --0.25&  0.10&    52&      --0.27&  0.12&    60&      --0.26&  0.51&    32&      --0.26&  0.16&    41&     --0.31&  0.11&    39&       --0.29&  0.11&    61     \\
Y {\sc ii }  &  --0.02&  0.10&     8&      --0.11&  0.15&     9&      --0.23&  0.14&     4&      --0.07&  0.19&     7&     --0.11&  0.13&    14&       --0.06&  0.12&     7     \\
Zr {\sc ii}  &  --0.07&  0.11&     2&      --0.04&  0.15&     3&      --0.11&  0.14&     2&      --0.15&  0.01&     2&     --0.23&  0.00&     1&       --0.17&  0.02&     2     \\
Ba {\sc ii}  &  {\bf0.13}&...&     3&       {\bf0.25}&...&   3&       {\bf0.21}&...&     3&           {\bf0.17}&...& 3&     {\bf0.05}&...&    3&         {\bf0.27}&...&    3     \\
La {\sc ii}  &    0.03&  0.07&     2&        0.06&  0.01&     2&      --0.01&  0.00&     1&        0.02&  0.00&     2&       0.04&  0.10&     2&       --0.03&  0.06&     2     \\
Ce {\sc ii}  &    0.12&  0.18&     7&      --0.08&  0.08&     9&        0.33&  0.16&     3&      --0.07&  0.15&     5&     --0.01&  0.26&     8&         0.00&  0.18&     8     \\
Nd {\sc ii}  &    0.04&  0.17&     9&        0.00&  0.14&    10&      --0.04&  0.14&     6&        0.18&  0.21&     9&     --0.05&  0.11&     7&       --0.08&  0.13&    10     \\
Eu {\sc ii}  &  {\bf0.10}&...&     1&      {\bf0.05}&...&     1&       {\bf0.25}&...&    1&      {\bf0.25}&...&     1&      {\bf0.15}&...&    1&       {\bf0.15}&...&     1     \\
\hline
\end{tabular}
  \\
Notes. The abundance values computed with synthetic spectrum marked as bold.
 \end{table*}

\begin{table*}
\caption{Abundance results for NGC 5822.}
\label{ngc5822}
\begin{tabular}{lrrrrrrrrrrrr}
\hline
& \multicolumn{3}{c}{1329} & \multicolumn{3}{c}{1645}&\multicolumn{3}{c}{1889}&\multicolumn{3}{c}{2397}\\
\hline
  Ion& [El/H]& $\sigma$ & NL &   [El/H] & $\sigma$ & NL &  [El/H] & $\sigma$ & NL &  [El/H] & $\sigma$ & NL\\
\hline
 O {\sc i}  &  {\bf0.20}&...&  1& {\bf0.15}& ...& 1& {\bf0.10}& ...&   1& {\bf0.20}&...&    1 \\
Na {\sc i}  &  {\bf0.11}&...&  4& {\bf0.09}& ...& 4& {\bf0.07}& ...&   4& {\bf0.04}&...&    4 \\
Mg {\sc i}  &  {\bf0.00}&...&  7& {\bf--0.03}& ...& 7& {\bf--0.03}& ...&   7& {\bf--0.08}&...&   7 \\
Al {\sc i}  & --0.03 & 0.01 &  2&--0.05&  0.08&  2&    0.03&  0.03&   2&  --0.05&  0.08&   2 \\
Si {\sc i}  &   0.07 & 0.13 & 22&  0.00&  0.11& 18&  --0.01&  0.13&  20&    0.02&  0.16&  22 \\
Ca {\sc i}  &   0.03 & 0.08 & 16&--0.02&  0.06& 16&    0.05&  0.09&  16&    0.04&  0.09&  17 \\
Sc {\sc ii} &   0.07 & 0.17 & 13&--0.03&  0.11& 10&  --0.12&  0.13&   8&    0.01&  0.14&  10 \\
Ti {\sc i } & --0.07 & 0.13 & 64&--0.13&  0.08& 53&  --0.13&  0.08&  50&  --0.09&  0.11&  55 \\
Ti {\sc ii} &   0.05 & 0.12 &  5&  0.03&  0.06&  5&    0.04&  0.14&   3&    0.06&  0.11&   4 \\
V {\sc i  } & --0.16 & 0.09 & 34&--0.20&  0.10& 33&  --0.18&  0.12&  33&  --0.20&  0.10&  35 \\
Cr {\sc i } & --0.09 & 0.12 & 37&--0.13&  0.09& 33&  --0.15&  0.12&  33&  --0.13&  0.09&  35 \\
Cr {\sc ii} &   0.11 & 0.14 &  2&  0.18&  0.07&  5&    0.14&  0.17&   5&    0.23&  0.10&   4 \\
Fe {\sc i } &   0.04 & 0.07 &211&--0.02&  0.09&242&  --0.02&  0.09& 235&    0.02&  0.09& 253 \\
Fe {\sc ii} &   0.02 & 0.05 & 23&--0.03&  0.13&  9&    0.00&  0.06&   8&    0.04&  0.08&   8 \\
Co {\sc i } & --0.02 & 0.12 & 26&--0.07&  0.13& 26&  --0.07&  0.10&  23&  --0.07&  0.13&  26 \\
Ni {\sc i } & --0.01 & 0.08 & 76&--0.05&  0.09& 76&  --0.06&  0.09&  72&  --0.04&  0.08&  72 \\
Y {\sc ii } &   0.22 & 0.10 &  9&  0.07&  0.09&  5&    0.13&  0.10&   6&    0.22&  0.15&   6 \\
Zr {\sc ii} &   0.13 & 0.08 &  4&  0.09&  0.14&  3&  --0.02&  0.15&   3&    0.09&  0.14&   3 \\
Ba {\sc ii} &  {\bf0.42}&...&  3& {\bf0.38}& ...& 3& {\bf0.36}& ...&   3& {\bf0.41}&...&   3 \\
La {\sc ii} &   0.23 & 0.06 &  2&  0.15&  0.05&  2&    0.10&  0.00&   1&    0.16&  0.03&   2 \\
Ce {\sc ii} &   0.15 & 0.06 &  8&  0.09&  0.06&  7&  --0.01&  0.11&   7&    0.08&  0.08&   7 \\
Nd {\sc ii} &   0.16 & 0.14 & 13&  0.09&  0.11& 11&  --0.01&  0.11&  10&    0.08&  0.13&  12 \\
Eu {\sc ii} &  {\bf0.15}&...&  1& {\bf0.10}& ...& 1& {\bf0.00}& ...&  1&  {\bf0.05}&...&   1 \\
\hline
\end{tabular}
  \\
Notes. The abundance values computed with synthetic spectrum marked as bold.
 \end{table*}

\begin{table*}
\caption{The mean elemental abundances in OCs.}
\label{mean}
\begin{tabular}{lrrrrrrrrrr}
\hline
 & \multicolumn{2}{c}{Cr 110} & \multicolumn{2}{c}{Cr 261}&\multicolumn{2}{c}{NGC 2477}  & \multicolumn{2}{c}{NGC 2506} & \multicolumn{2}{c}{NGC 5822} \\
\hline
  Ion& [El/H]& $\sigma$ & [El/H] & $\sigma$ &  [El/H] & $\sigma$ &  [El/H] & $\sigma$ &[El/H] & $\sigma$  \\
\hline
 O {\sc i}   &  0.05&... &  ... &...  &--0.13  & ... &--0.06 & ... &  0.17 & ...  \\
Na {\sc i}   &--0.02&... &  0.13&...  &  0.12  & ... &--0.14 & ... &  0.07 & ...  \\
Mg {\sc i}   &--0.07&... &  0.04&...  &--0.03  & ... &--0.22 & ... &--0.04 & ...  \\
Al {\sc i}   &  0.04&0.04&  0.10&0.09 &--0.08  &0.08 &--0.02 &0.05 &--0.02 &0.04  \\
Si {\sc i}   &  0.03&0.12&  0.05&0.13 &  0.17  &0.16 &--0.18 &0.12 &  0.03 &0.14  \\
Ca {\sc i}   &--0.02&0.11&--0.03&0.11 &  0.07  &0.14 &--0.17 &0.09 &  0.04 &0.09  \\
Sc {\sc ii}  &--0.03&0.14&  0.08&0.12 &  ...   &...  &--0.10 &0.13 &  0.00 &0.15  \\
Ti {\sc i }  &--0.04&0.10&  0.04&0.13 &--0.01  &0.08 &--0.25 &0.12 &--0.09 &0.11  \\
Ti {\sc ii}  &--0.00&0.11&  0.07&0.07 &  0.11  &0.15 &--0.05 &0.06 &  0.05 &0.12  \\
V {\sc i  }  &--0.05&0.10&  0.09&0.12 &  0.07  &0.13 &--0.24 &0.10 &--0.18 &0.10  \\
Cr {\sc i }  &--0.01&0.09&  0.01&0.10 &  0.04  &0.12 &--0.28 &0.13 &--0.12 &0.11  \\
Fe {\sc i }  &--0.02&0.11&--0.01&0.11 &  0.17  &0.11 &--0.22 &0.12 &  0.01 &0.08  \\
Fe {\sc ii}  &--0.10&0.09&--0.07&0.09 &  0.13  &0.14 &--0.27 &0.12 &  0.02 &0.06  \\
Co {\sc i }  &--0.05&0.09&  0.03&0.16 &  0.14  &0.11 &--0.26 &0.14 &--0.05 &0.12  \\
Ni {\sc i }  &--0.06&0.10&  0.04&0.11 &  0.12  &0.10 &--0.28 &0.12 &--0.04 &0.08  \\
Y {\sc ii }  &  0.06&0.12&  0.07&0.18 &  0.11  &0.15 &--0.08 &0.14 &  0.19 &0.11  \\
Zr {\sc ii}  &  0.03&0.13&  0.06&0.13 &  0.15  &0.11 &--0.11 &0.07 &  0.07 &0.12  \\
Ba {\sc ii}  &  0.32& ...&  0.32& ... &  0.28  & ... &  0.17 & ... &  0.40 & ...  \\
La {\sc ii}  &  0.15&0.08&  0.14&0.06 &  0.23  &0.03 &  0.02 &0.05 &  0.18 &0.04  \\
Ce {\sc ii}  &  0.12&0.14&  0.06&0.16 &  0.16  &0.16 & -0.01 &0.17 &  0.08 &0.08  \\
Nd {\sc ii}  &  0.08&0.11&  0.06&0.14 &  0.07  &0.15 &  0.02 &0.15 &  0.08 &0.13  \\
Eu {\sc ii}  &  0.20& ...&  0.25& ... &  0.17  & ... &  0.14 & ... &  0.07 & ...  \\
\hline
\end{tabular}
\end{table*}

\begin{table}
\caption{Comparison of our results for star 2129 in  Cr~110 with
 that obtained by Pancino et al. (2010).}
\label{compar}
\begin{tabular}{lrrrrr}
\hline
  Ion& [El/H]  & $\sigma$ & NL &   [El/H]   & $\sigma$   \\
     &  (our)  &          &    &  (Pan2010)  &            \\
\hline
 O {\sc i}   &       {\bf0.05}& --&   1&      --0.02&  0.12  \\
Na {\sc i}   &      {\bf--0.02}& --&  4&      --0.01&  0.08   \\
Mg {\sc i}   &     {\bf--0.07}& --&   7&        0.06&  0.14   \\
Al {\sc i}   &       0.10&  0.05&     2&        0.01&  0.08  \\
Si {\sc i}   &       0.00&  0.13&    17&        0.09&  0.02  \\
Ca {\sc i}   &     --0.06&  0.11&    12&        0.01&  0.04  \\
Sc {\sc ii}  &     --0.09&  0.07&     8&      --0.02&  0.06  \\
Ti {\sc i }  &     --0.04&  0.08&    25&        0.05&  0.03   \\
Ti {\sc ii}  &     --0.09&  0.14&     4&      --0.04&  0.07   \\
V {\sc i  }  &     --0.03&  0.09&    19&        0.02&  0.05   \\
Cr {\sc i }  &     --0.01&  0.06&     7&        0.01&  0.06  \\
Fe {\sc i }  &     --0.04&  0.09&   119&        0.05&  0.01   \\
Fe {\sc ii}  &     --0.07&  0.07&     7&      --0.04&  0.08   \\
Co {\sc i }  &     --0.05&  0.10&    13&      --0.03&  0.04   \\
Ni {\sc i }  &     --0.09&  0.07&    37&      --0.06&  0.02   \\
Y {\sc ii }  &       0.01&  0.14&     8&      --0.12&  0.08   \\
Zr {\sc ii}  &     --0.09&  0.18&     3&        0.00&  0.15   \\
Ba {\sc ii}  &      {\bf0.31}& --&    3&        0.54&  0.04  \\
La {\sc ii}  &       0.10&  0.05&     3&        0.12&  0.03   \\
Ce {\sc ii}  &       0.09&  0.09&     3&        0.02&  0.12  \\
Nd {\sc ii}  &       0.03&  0.11&     9&        0.29&  0.13   \\
Eu {\sc ii}  &     {\bf0.16}&...&     1&         ...&  ...   \\
\hline
\end{tabular}
 \end{table}

\begin{table}
\caption{Comparison of our results for star 3231 in  NGC~2506 with that
obtained by Reddy et al. (2012).}
\label{compar2}
\begin{tabular}{lrrrrrr}
\hline
  Ion& [El/H]     & $\sigma$ & NL &   [El/H]    & $\sigma$ & NL \\
  Ion&   (our)    &          &    & (Reddy2012)   &          &    \\
\hline
 O {\sc i}   &      {\bf--0.10}& --&   1&       --0.19&  ... &    1  \\
Na {\sc i}   &      {\bf--0.16}& --&   4&       --0.11&  0.08&    5  \\
Mg {\sc i}   &      {\bf--0.23}& --&   7&       --0.22&  0.07&    3   \\
Al {\sc i}   &      --0.03&  0.03&     2&       --0.06&  0.03&    2  \\
Si {\sc i}   &      --0.19&  0.13&    19&       --0.22&  0.08&    7   \\
Ca {\sc i}   &      --0.13&  0.08&    12&       --0.16&  0.09&    9  \\
Sc {\sc ii}  &      --0.07&  0.06&    11&       --0.16&  0.09&    5  \\
Ti {\sc i }  &      --0.24&  0.12&    33&       --0.26&  0.09&    9 \\
Ti {\sc ii}  &        0.02&  0.04&     2&       --0.12&  0.06&    6 \\
V {\sc i  }  &      --0.23&  0.14&    19&       --0.20&  0.07&    8  \\
Cr {\sc i }  &      --0.31&  0.13&    11&       --0.27&  0.12&    9  \\
Fe {\sc i }  &      --0.22&  0.13&   121&       --0.25&  0.06&   38  \\
Fe {\sc ii}  &      --0.28&  0.11&     9&       --0.22&  0.06&    8  \\
Co {\sc i }  &      --0.27&  0.14&    18&       --0.30&  0.14&    26 \\
Ni {\sc i }  &      --0.31&  0.11&    39&       --0.34&  0.11&    61 \\
Y {\sc ii }  &      --0.11&  0.13&    14&       --0.22&  0.12&    1  \\
Zr {\sc ii}  &      --0.23&  0.00&     1&        ...  &   ...&   ... \\
Ba {\sc ii}  &       {\bf0.05}& --&   3&          0.06&   ...&   ... \\
La {\sc ii}  &        0.04&  0.10&     2&         0.06&  0.07&    1  \\
Ce {\sc ii}  &      --0.01&  0.26&     8&        ...  &   ...&  ... \\
Nd {\sc ii}  &      --0.05&  0.11&     7&       --0.08&  0.13&    10\\
Eu {\sc ii}  &       {\bf0.15}& --&    1&         0.01&   ...&  ... \\
\hline
\end{tabular}
 \end{table}

\begin{table}
\caption{Comparison of our results for stars 2291 and 2311  in  Cr~261
with that obtained by de Silva et al. (2007).}
\label{compar3}
\begin{tabular}{lrrrrrr}
\hline
    &              2291 &&                    2311 &&     \\
\hline
  Ion        & log A  &    log A  &  log A      &    log A  \\
     &       (our) &   (DS2007) &        (our)      &     (DS2007)  \\
\hline
Na {\sc i } &  6.44  &      6.45 &        6.39 &       6.65  \\
Mg {\sc i } &  7.61  &      7.67 &        7.62 &       7.89  \\
Si {\sc i } &  7.60  &      7.66 &        7.60 &       7.85  \\
Ca {\sc i } &  6.29  &      6.29 &        6.38 &       6.61  \\
Fe {\sc i } &  7.57  &      7.51 &        7.55 &       7.56  \\
Fe {\sc ii} &  7.53  &       ... &        7.48 &        ...  \\
Ni {\sc i } &  6.31  &      6.19 &        6.37 &       6.33  \\
Ba {\sc ii} &  2.57  &      2.13 &        2.53 &       2.37  \\
\hline
\end{tabular}
\end{table}

\begin{figure} 
\begin{tabular}{c}
\includegraphics[width=8cm]{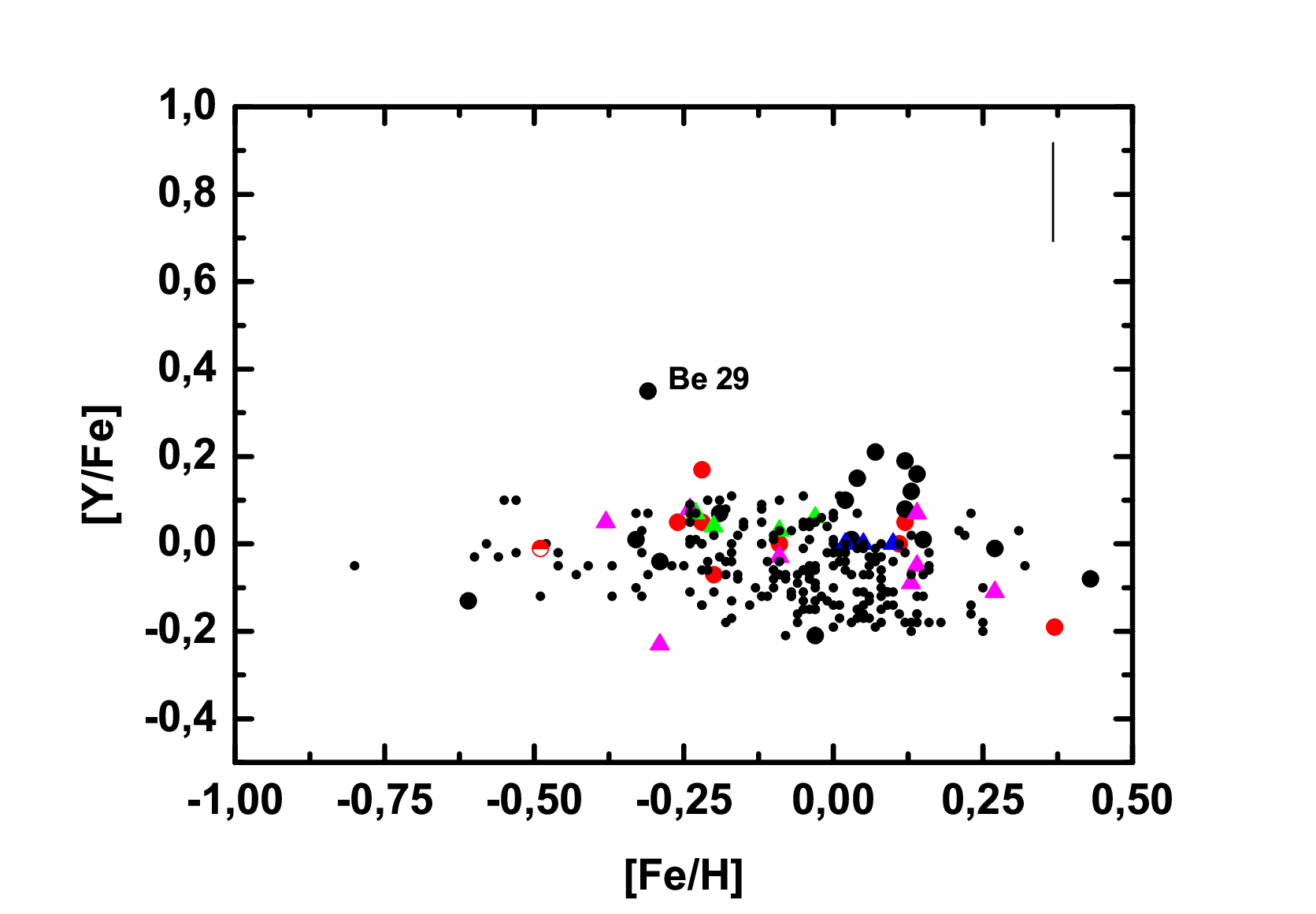}\\
\includegraphics[width=8cm]{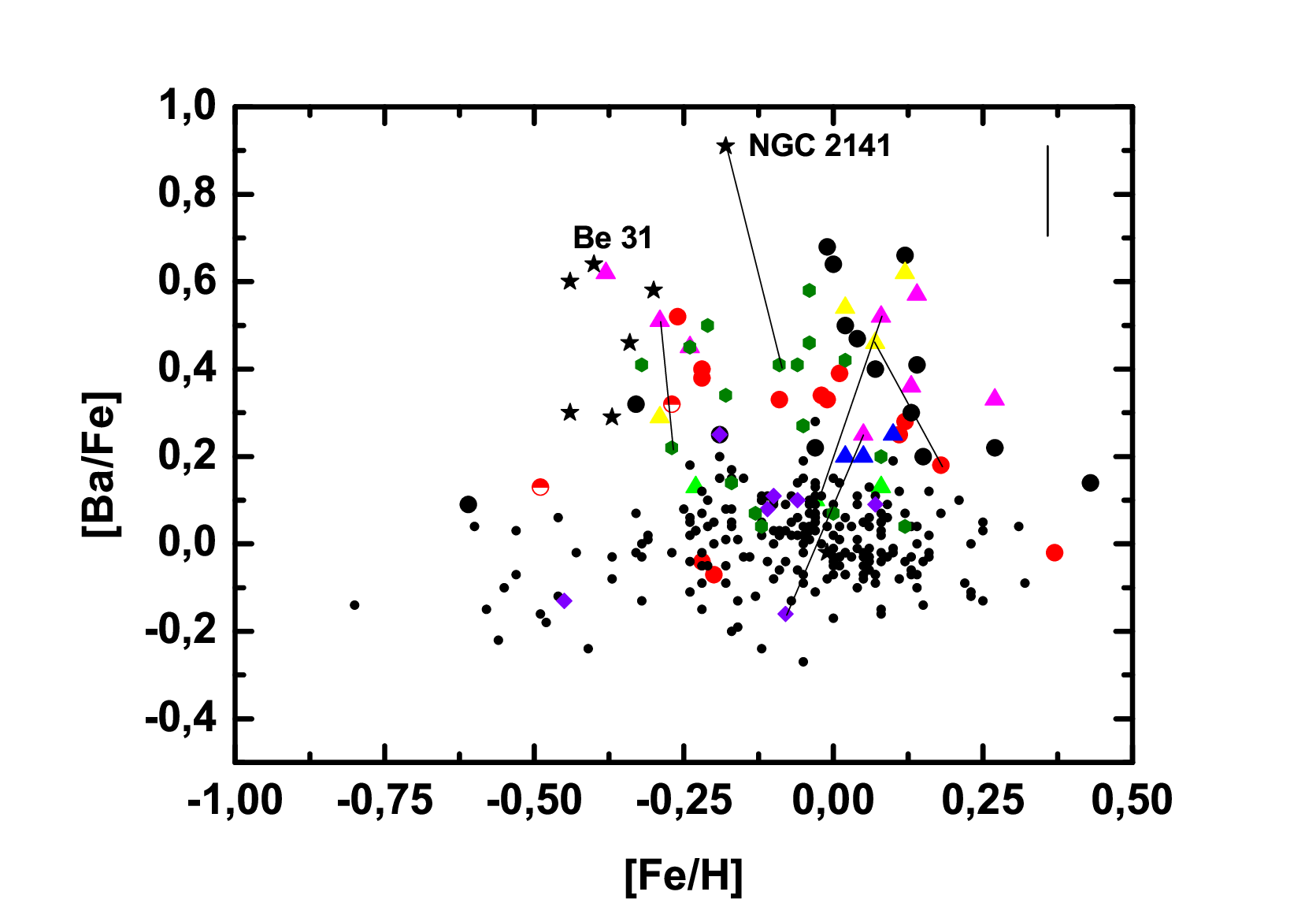}\\
\includegraphics[width=8cm]{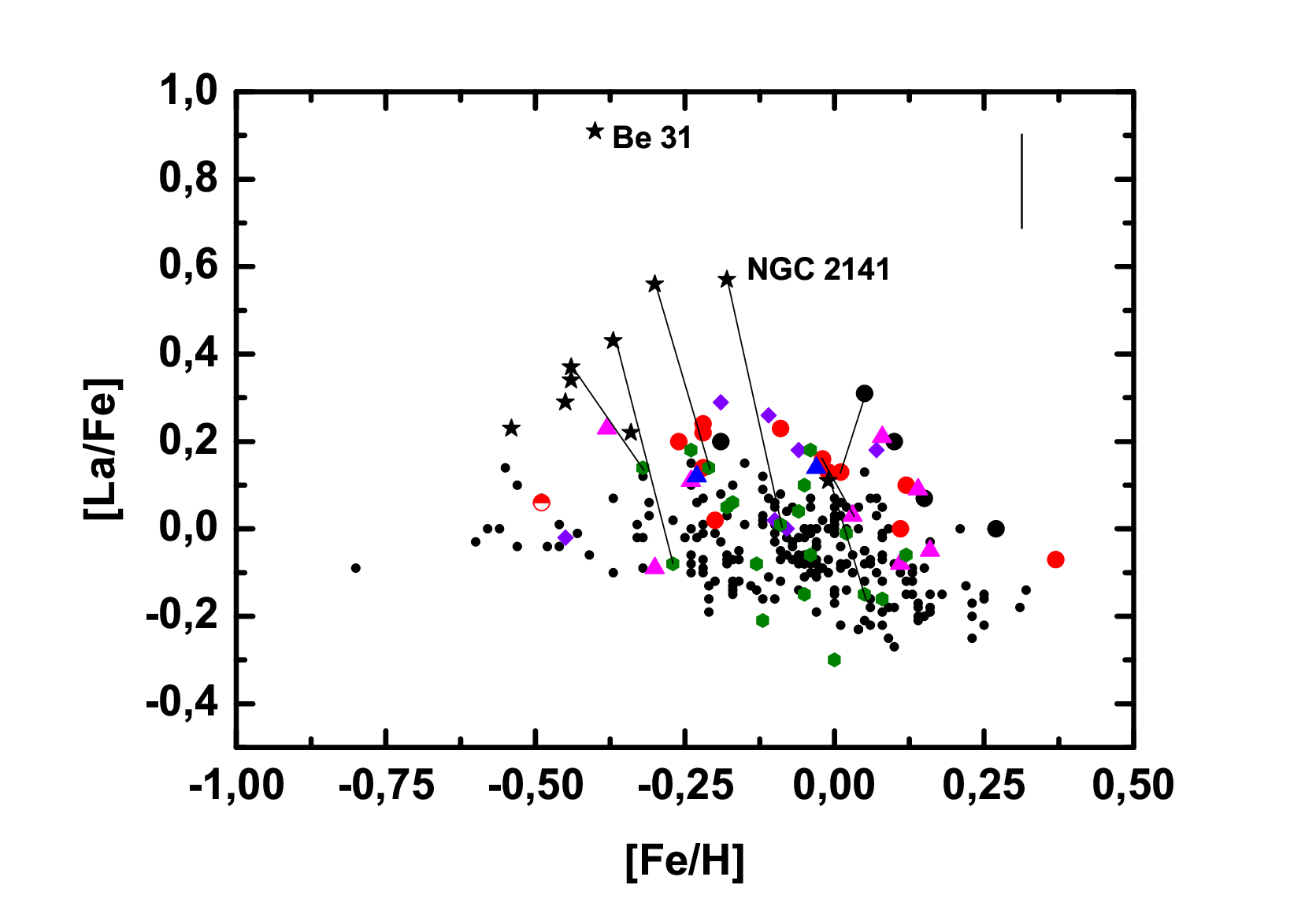}\\
\end{tabular}
\caption{
The trend of [Y/Fe] (upper panel), [Ba/Fe] (central
panel) and [La/Fe] (lower panel) versus [Fe/H]. Symbols are as fol-
lows: Y and La abundances by Maiorca et al. (2011) and Ba abun-
dances by D'Orazi et al. (2009): black circles; Pancino et al.(2010)
and Carrera \& Pancino (2011): magenta triangles; D'Orazi et al.
(2012) : blue triangles; Reddy et al. (2012) : green triangles.
Ba abundances by Bragaglia et al. (2008): yellow
triangles; Yong et al. (2005. 2012) : asterisks; Reddy et al. (2013) :
violet rhombuses; Jacobson \& Friel (2013) : olive diamonds; Car-
raro et al. (2014) and Monaco et al. (2014) : semi-full red cirles;
Mishenina et al. (2013a) : the thin disc (marked as black dots);
finally, our determinations (Mishenina et al. 2013b, Mishenina et
al. 2014) and in present study : red circles.
}
\label{el_fe}
\end{figure}

\section{Results and comparison with the literature}

The chemical composition for our program  clusters was the subject of
several studies in the past.
The purpose of this study, however, is mostly to analyze the behavior
of the neutron-capture elements.
Here we summarise previous measurements and compare with our results.
A detailed comparison between this work and the literature is given
in Table~\ref{comparison_literature}. \\

\begin{table*}
\caption{Comparison of the data obtained in the works of various authors.}
\label{comparison_literature}
\begin{tabular}{lrrrrl}
\hline
         & [Fe/H] & [Ba/Fe]& [Y/Fe] & [La/Fe] &  Ref.    \\
\hline
Cr 110   & --0.02 &  0.34 &   0.08 &   0.16  &  this work      \\
          &   0.03 &  0.49 &   0.10 &   0.03  &  \cite{pan10}   \\
Cr 261   & --0.01 &  0.33 &   0.07 &   0.13  &  this work      \\
          & --0.03 &  0.30 &   ...  &   ...   &  \cite{car05}   \\
          & --0.03 &  0.03 &   ...  &   ...   &  \cite{sil07}*   \\
          &   0.13 &  ...  & --0.21 &   ...   &  \cite{mai11}   \\
          &   0.13 &  0.22 &   ...  &   ...   &  \cite{dor09}   \\
NGC 752   &   0.01 &  0.19 &   ...  &   ...   &  \cite{dor09}   \\
          & --0.02 &  0.13 &   0.04 &   ...   &  \cite{red12}   \\
          &   0.08 &  0.52 & --0.03 &  0.18   &  \cite{car11}   \\
NGC 2141  & --0.09 &  0.41 &   ...  &  0.01   &  \cite{jfr13}   \\
          & --0.18 &  0.91 &   ...  &  0.57   &  \cite{yon05}   \\
NGC 2477  &   0.18 &  0.18 &   0.12 &  0.16   &  this work      \\
          &   0.07 &  0.46 &   ...  &   ...   &  \cite{bra08}   \\
          &   0.07 &  ...  &   0.21 &   ...   &  \cite{mai11}   \\
NGC 2506  & --0.22 &  0.40 &   0.12 &   0.24  &  this work      \\
          & --0.19 &  0.31 &   0.04 &   0.28  &  \cite{red12}   \\
          & --0.24 &  0.04 &   ...  &   ...   &  \cite{mik12}*  \\
NGC 2660  &   0.04 &  0.47 &   ...  &   ...   &  \cite{dor09}   \\
          &   0.04 &  0.61 &   ...  &   ...   &  \cite{bra08}   \\
          &   0.04 &  ...  &    0.15&   ...   &   \cite{mai11}  \\
NGC 5822  &   0.01 &  0.39 &   0.12 &   0.13  &  this work      \\
          &   0.05 &  ...  &   ...  &    0.31 &   \cite{mai11}  \\
Be 18     & --0.44&  0.30&   ...  &    0.34 &   \cite{yon12}  \\
          & --0.32 &  0.41&  ... &    0.14  &   \cite{jfr13}    \\
Be 20     & --0.45 &  0.14 &   ...  &   0.30  &  \cite{yon05}   \\
          & --0.30 &  0.09 &   ...  &   ...   &  \cite{dor09}   \\
          & --0.30 &  ...  &  --0.13&   ...   &  \cite{mai11}   \\
Be 21     &--0.30 & 0.58&   ... &    0.56  &  \cite{yon12}   \\
          & --0.21 &  0.50&  ... &    0.14  &   \cite{jfr13}    \\
Be  22    & --0.44 &  0.60&   ... &    0.37  &    \cite{yon12}     \\
          & --0.24 &  0.45&  ... &    0.18  &   \cite{jfr13}    \\
Be 29     & --0.31 &  0.40 &   ...  &   ...   &  \cite{dor09}   \\
          & --0.54 &  0.30 &   ...  &   ...   &  \cite{yon05}   \\
          & --0.31 &  ...  &   0.35 &   ...   &  \cite{mai11}   \\
Be 32     & --0.30 &  0.51 &  --0.23&  --0.14 &  \cite{car11}   \\
          & --0.29 &  0.24 &   ...  &   ...   &  \cite{dor09}   \\
          & --0.29 &  ...  &  --0.04&   ...   &  \cite{mai11}   \\
          & --0.37 &  0.29&   ... &    0.43  &    \cite{yon12}   \\
          & --0.27 &  0.22 &   ...  &  --0.08 &  \cite{jfr13}   \\
          & --0.29 &  0.29 &   ...  &   ...   &  \cite{bra08}   \\
Hyades    &   0.11 &  0.36 &  --0.09&  --0.08 &  \cite{car11}   \\
          &   0.13 &  0.30 &   ...  &   ...   &  \cite{dor09}   \\
          &   0.13 &  ...  &   0.12 &   ...   &  \cite{mai11}   \\
Praesepe  &   0.16 &  0.33 &  --0.11&  --0.05 &  \cite{car11} \\
          &   0.27 &  0.22 &   ...  &   ...   &  \cite{dor09}   \\
          &   0.27 &  ...  &   -0.01&   ...   &  \cite{mai11}   \\
M 67      &   0.03 &  ...  &   0.01 &   0.06  &  \cite{mai11}   \\
          &   0.05 &  0.25 & --0.05 &   0.05  &  \cite{pan10}   \\
          &   0.02 &  0.04 &   ...  &   ...   &  \cite{dor09}   \\
          &   0.05 &  0.10 &   ...  &  --0.15 &  \cite{jfr13}   \\
          & --0.01 &--0.02 &   ...  &    0.11 &  \cite{yon05}   \\
          & --0.08 &--0.16 &  0.03  &    0.00 &  \cite{red13}   \\
PWM4      & --0.34 &  0.36 &   ...  &    0.22 &  \cite{yon12}   \\
          & --0.18 &  0.34 &   ...  &    0.05 &  \cite{jfr13}   \\
\hline
\end{tabular}
\\
 Notes: * -- These data are not all included in the figures. See the text.
\end{table*}

\noindent
{\bf Collinder 110.} \\
This cluster has significant reddening (E(B--V)=0.54$\pm$0.03 \citep{pan10}.
\cite{pan10} obtained accurate abundances of seventeen elements, included Y, Ba, La, and Nd.
With a cluster metallicity [Fe/H] = +0.03$\pm$0.10, they found a significant
barium overabundance ([Ba/Fe] = 0.49 $\pm$0.06), and excess of
neodymium [Nd/Fe] = 0.23$\pm$0.20.
The values of the yttrium [Y/Fe] = --0.10$\pm$0.12 and lanthanum
[La/Fe] = +0.03$\pm$0.18 are instead close to solar.

We obtained the mean values of [Fe/H] = --0.02, a
moderate excess of [Ba/Fe] = 0.34 and a slight excess
of [Y/Fe] = 0.08 and [La/Fe] = 0.16.\\

\noindent
{\bf Collinder 261}. \\
The reddening of this cluster has been derived several times:
E(B--V) is about  0.22 (the value is quite uncertain, 
\citep{maz95},  0.33 \citep{jan94},   0.25 --0.34 \citep{goz96}.
The same is true
for its chemical composition, which, however,  shows
significant study-to-study variations:
[Fe/H]= --0.16  \citep{fr02},  --0.22  \citep{fr03},  --0.03  \citep{car05},
--0.03 \citep{sil07},  +0.13 \citep{ses08}, 0.00 \citep{mik12}.
Concerning neutron-capture elements,
a moderate excess of  barium [Ba/Fe]  = 0.30$\pm$0.08 was found by \cite{car05}, while
a sub-solar values of [Zr/Fe] =  0.12  and [Ba/Fe] = 0.03 with an
intrinsic scatter smaller than 0.05 dex  were derived by \cite{sil07}.

These results are consistent with \cite{car05}, while there is a
discrepancy of about 0.3 dex with the [Ba/Fe] calculated by \cite{sil07}.
[Y/Fe] = --0.21$\pm$0.07 was found by \cite{mai11}.

We derived a mean values of [Fe/H] = --0.01,
a moderate excess of [Ba/Fe] = 0.33,  and a slight excess of [Y/Fe] = 0.07
and [La/Fe] = +0.13.\\

\noindent
{\bf NGC 2477.}\\
This cluster has an average reddening E(B--V) = 0.29 \citep{har72},
and more recent estimates confirm this early result.
\cite{bra08} determined a metallicity  [Fe/H] = +0.07 $\pm$0.03 and
[Ba/Fe] = 0.46$\pm$0.05.
[Y/Fe] = 0.21$\pm$0.09 was found by \cite{mai11}.

In our case,
we obtained [Fe/H] = +0.15, and  we detected only a slight excess of
[Ba/Fe] = 0.15, while [Y/Fe] = --0.05 and [La/Fe] = 0.08 are close to solar.
In particular, the [Ba/Fe] that we calculated is about 0.3 dex lower
than \cite{bra08}.

C14 derived a mean iron content of [Fe/H]=0.09, or [Fe/H]=0.04
adopting the same solar iron content adopted here.\\

\noindent
{\bf NGC 2506.}\\
E(B--V) is the range 0.0--0.07 \citep{mar97}.
Several estimates of iron abundance are available:
[Fe/H] = --0.44 $\pm$0.06  \citep{fr02},   [Fe/H] = --0.20$\pm$0.02 (from 2 stars, \cite{car04}),
[Fe/H] =--0.19$\pm$0.06    \citep{red12},  [Fe/H] = --0.24$\pm$0.05 \citep{mik12}.
\cite{red12} provided the following estimates for
$n$-capture element abundance: [Y/Fe] = 0.04 $\pm$0.07, [Ba/Fe] = 0.31,
[La/Fe] = 0.28$\pm$0.4, [Ce/Fe] = 0.18,  [Nd/Fe] = 0.16$\pm$0.06,
[Sm/Fe] = 0.22, and [Eu/Fe] = 0.22 . On the other hand, \cite{mik12}
provided [Ba/Fe] = 0.04$\pm$0.10 and [Eu/Fe] = 0.20$\pm$0.03. 

Our analysis yields a mean values of [Fe/H] = --0.22, an excess of [Ba/Fe] = 0.40,
and [Y/Fe] = 0.12, and [La/Fe] = 0.24.
Our [Ba/Fe] is about 0.1 dex higher than Reddy et al.,
and almost 0.4 dex higher
than \cite{mik12}.\\

\noindent
{\bf NGC 5822.}\\
The value of  E(B--V) is in the range 0.10 --0.15 \citep{car11}, while
metallicity is measured as [Fe/H] = 0.04 \citep{smi09},
[Fe/H] = 0.05 \citep{pac10}, and  [Fe/H]  = --0.058$\pm$0.027  \citep{car11}.
[La/Fe] = 0.31$\pm$0.01 was found by \cite{mai11}.

We obtained the mean values of [Fe/H] = 0.01, an excess of
[Ba/Fe] = 0.39 and lower excesses for Y and La, with
[Y/Fe] = 0.12 and [La/Fe] = 0.13. C14 derived a mean iron
content of [Fe/H] = 0.02, or [Fe/H] = --0.03 for
the same solar iron content adopted here.

\section{Results and discussion}

The main result from previous works is that [Ba/Fe] is larger than solar
for a number of OCs.
In particular, the [Ba/Fe] spread tends to increase with decreasing the
OCs age, 
with younger associations 
showing the largest overabundances
\citep[][, etc]{dor09,yon12,jfr13}.
More in general, OCs show a
larger spread of Ba enrichment compared to disk stars with similar
age (Mishenina et al. 2013ab, 2014).
We compared our findings with the results of other authors (Tables 12--14),
as well as data obtained in other studies (Table 15).
While for a number of OCs a good agreement is obtained, within the
observational errors, for other cases a significant departure is
observed in the results by different authors. This variation is
due to a number of reasons, including e.g., the quality and methods
of processing the spectra, atmospheric parameters, the used  the
atomic parameters, especially the oscillator strengths and damping
constants, physical approaches LTE or NLTE,  model atmospheres and
code abundance computations. This issue was extensively discussed
in previous works \citep[e.g.,][]{fr10, yon12}:
the lack of an homogeneous analysis and systematic abundance
differences can be much larger than the expected observational
errors. For the discussion in this section, we use the data of
other authors in their original form, without any correction.
Indeed, it is difficult to determine the cause of the difference
case by case. On the other hand, we will discuss the larger discrepancies.

Within the uncertainties, the Y enrichment is consistent with disk
stars, and consistent with the Sun within 0.2 dex. 
Therefore, it does not seem that there is any significant anomaly in
the Y abundance in OCs \citep[][]{pan10,mai11,mis13b}.

The situation is partially different for La. While a number of OCs are
consistent with the average of the stars in the disk, 
a significant fraction show a [La/Fe] about 0.2--0.3 dex larger than
in the Sun. These departures are beyond the present error estimations,
but they could be explained within the present systematic uncertanties
highlighted comparing the results from different authors
\citep[e.g.,][and references therein]{yon12,jfr13}.

On the other hand, as discussed in \cite{mis13b} for thin disk stars,
the interpretation of the trend of neutron-capture elements with
respect to Fe needs to also take into account that Fe is not a fully
primary element at high metallicities. 
In particular, the production
of Fe in thermonuclear supernovae
\citep[SNIa,][and references therein]{nomoto:13} is decreasing with 
increasing initial metallicity of the SNIa progenitor
\citep[e.g.,][]{timmes:03,travaglio:05,bravo:10}. This theoretical
prediction is confirmed by the observation of the
[Ni/Fe] increasing trend for super-solar thin disk stars, where
the bulk of Ni is instead fully primary \citep[see discussion in][]{mis13a}.
Therefore, with respect to the Sun a scatter of neutron-capture elements
compared to Fe may be expected in the disk and in OCs, depending on the
Fe enrichment history. The quantification of this intrinsic scatter due
to the Fe production from SNIa needs to be estimated by galactical chemical
evolution simulations, that take into account present uncertainties affecting
theoretical SNIa yields.

We cannot exclude that our sample is affected by
observational issues, especially for La. In Table \ref{errors} we
have shown that the expected uncertainty for the [La/Fe] is about 0.1
dex. On the other hand, there are much larger differences for La
between different works (e.g., Jacobson \& Friel 2013). 
Among others, there is the example of NGC~2141.
Be 31 and NGC~2141  show a [La/Fe] that is much
larger than other observed OCs: [La/Fe]= 0.91 and 0.57 \citep{yon05}.
The same OCs show extremely high [Ba/Fe] = 0.64 (Be 31) and 0.91 (NGC~2141),
and for  [Eu/Fe] = 0.56 (Be 31) and 0.17 (NGC~2141).
At the end of this section we will discuss again these two special cases.

For the element Eu, considered as a typical $r$-process element,
we found abundances consistent with the solar system and with
the average of the disk.

In order to derive additional observational constrains for stellar simulations and the chemical enrichment history of OCs, we combined here the results of our
analysis with data collected from the literature, and build up the largest sample to date of high resolution abundances of neutron-capture elements.

In Fig.~3 we show the trend of [Ba/Fe], [La/Fe] and [Y/Fe] versus
[Fe/H] for all the OCs available, and we include also Melotte~66
(Carraro et al. 2014),
and Trumpler~5 (Monaco et al. 2014).
In the figures, the abundance values obtained for the same
cluster and having the difference between these values more than
the errors given in Table 4 are connected by a line.

In Fig.~4 The [Ba/Fe], [La/Fe] and [Y/Fe] are also shown as a function of
the  cluster ages. The ages were calculated consistently,
according to \citet{car94}.
In the figures we include abundances obtained earlier (\citealt{mis13b,mis14})
with measurements from other authors, for a number of OCs
\citep[][]{dor09,pan10,car11,mai11,dor12,red12,red13,bra08,yon05,jfr13}
and the data for the thin disk stars were taken from the study by
\cite{mis13a}.

\begin{figure}
\begin{tabular}{c}
\includegraphics[width=8cm]{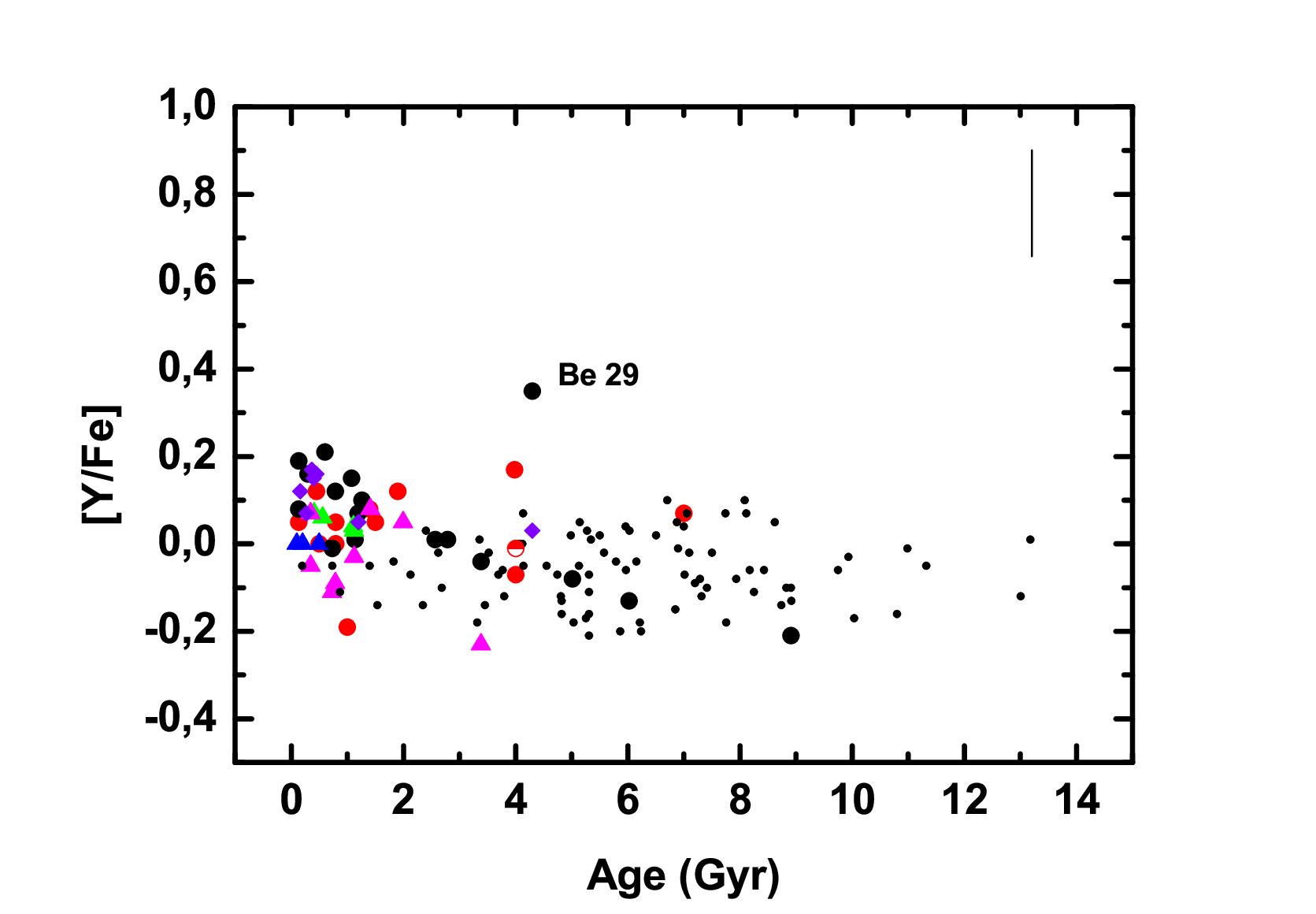}\\ \includegraphics[width=8cm]{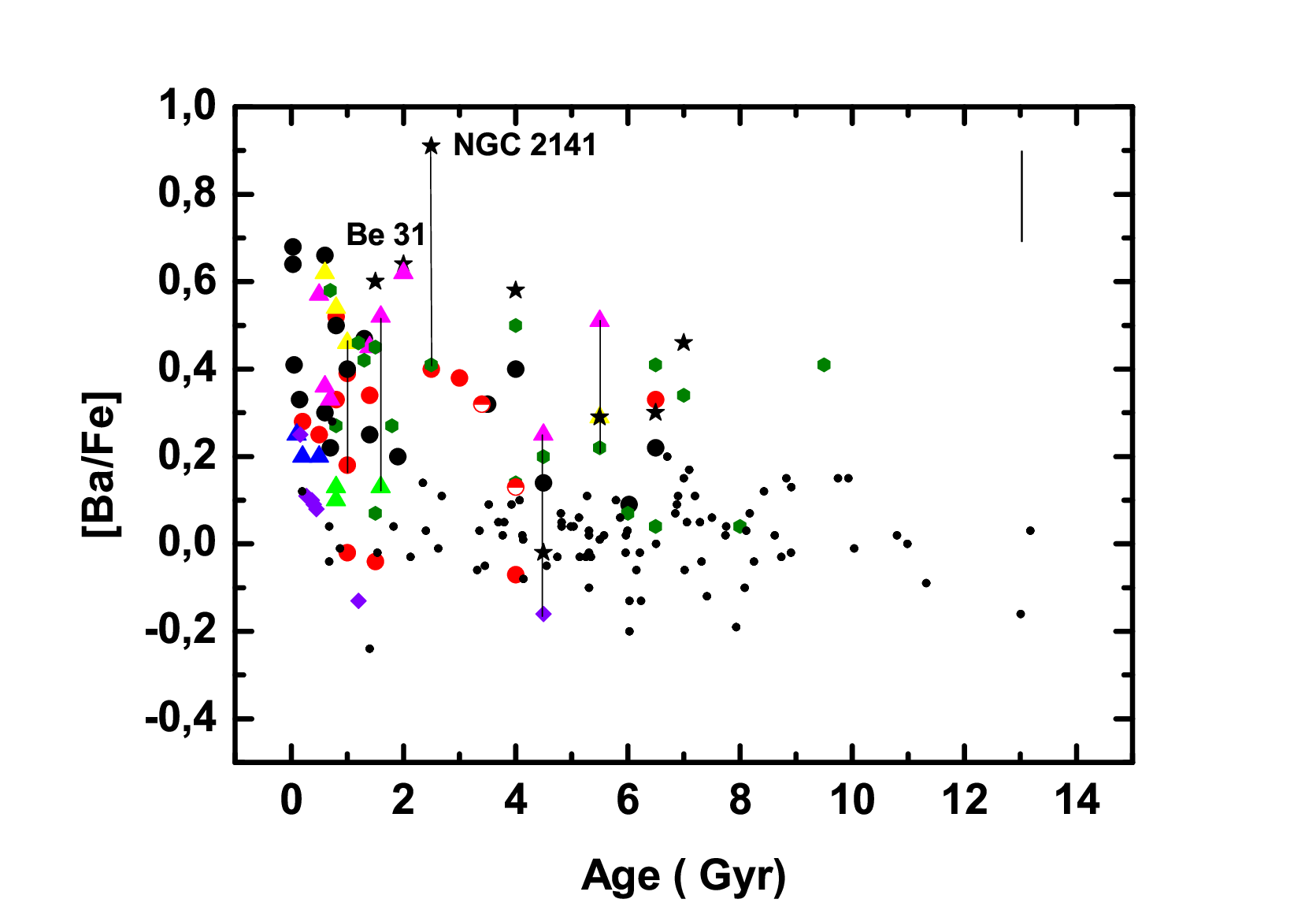}\\ \includegraphics[width=8cm]{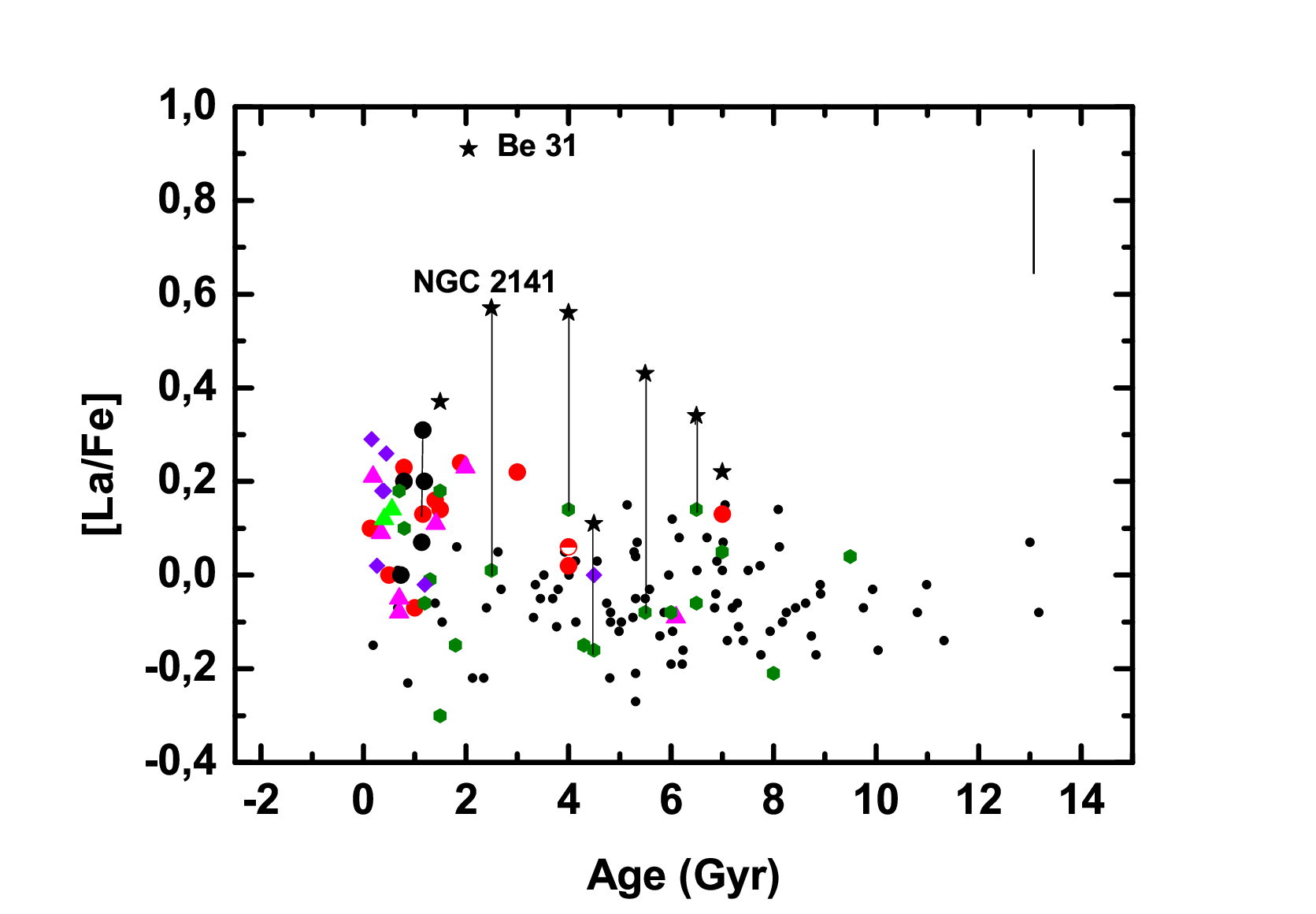}\\
\end{tabular}
\caption{The trend of [Y/Fe], [La/Fe] and [Ba/Fe] are reported compared
 to the Age. The age values were obtained in the uniform scale as
in Carraro \& Chiosi (1994).
 Symbols for different observations are reported as in Fig.~\ref{el_fe}.
  }
\label{el_time}
\end{figure}

\begin{figure}
\begin{tabular}{c}
\includegraphics[width=8cm]{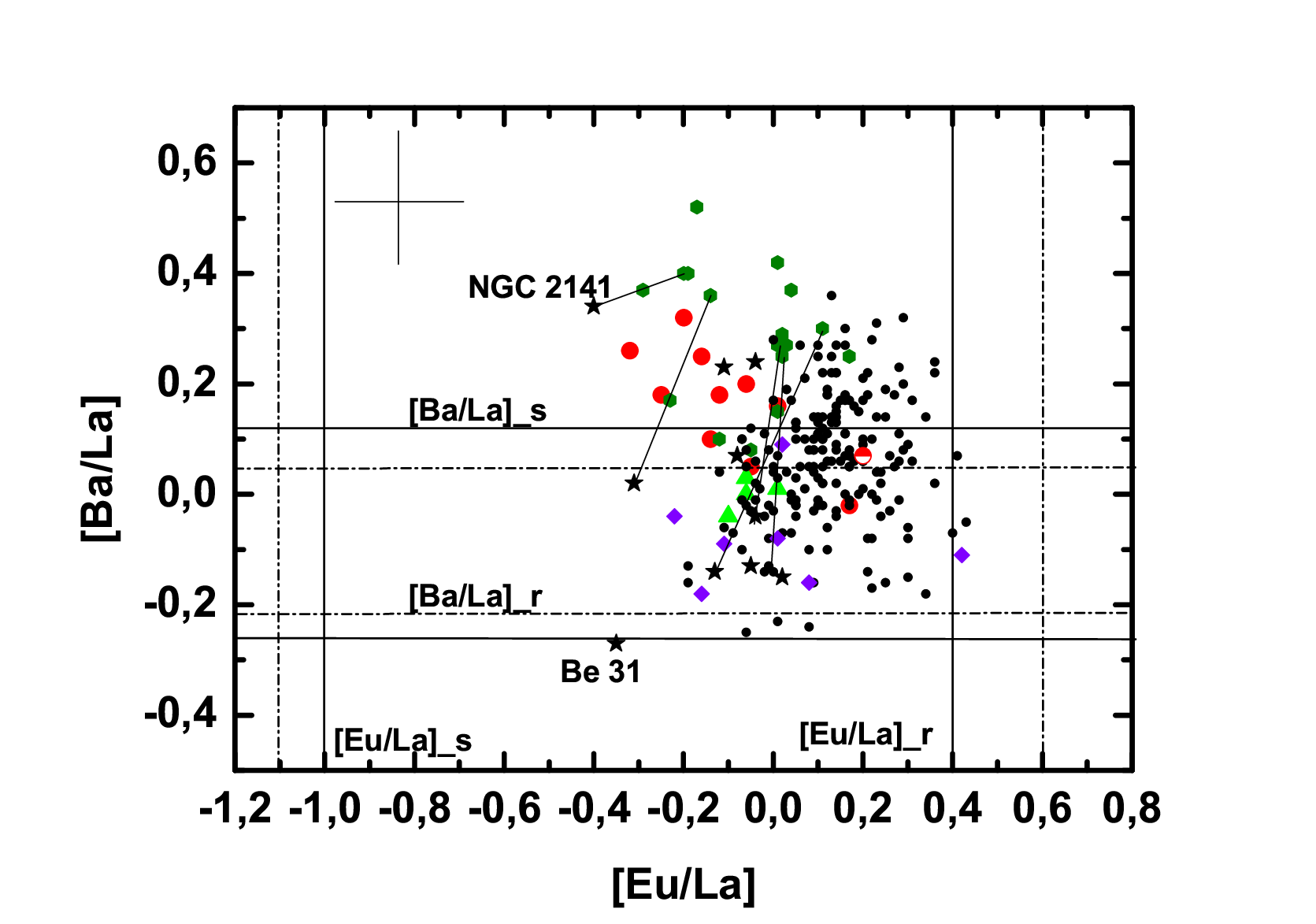}\\ \includegraphics[width=8cm]{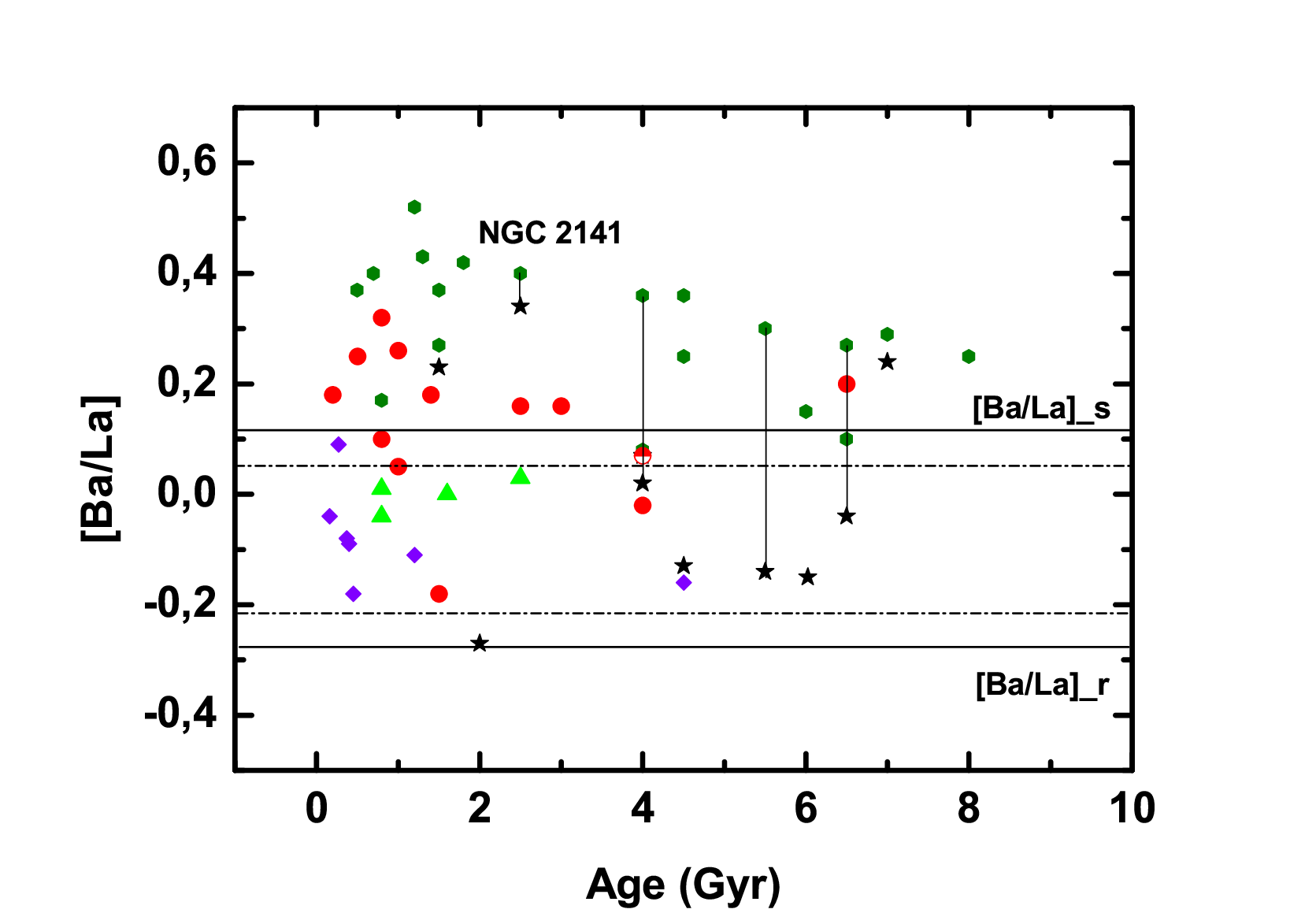}\\
\end{tabular}
\caption[]{{\it Upper panel}: The [Ba/La] ratio for a sample of OCs and disk stars
 is plotted with respect to the [Eu/La]. The pure $s$-process
 and $r$-process ratios are indicated in the figure, according
 to \cite{bisterzo:14} (dotted lines) and \cite{travaglio:04} (solid lines).
 {\it Bottom panel}: The [Ba/La] ratio vs. Age for a sample of OCs.
}
\label{bala_eula}
\end{figure}

\begin{figure}
\resizebox{\hsize}{!}
{\includegraphics{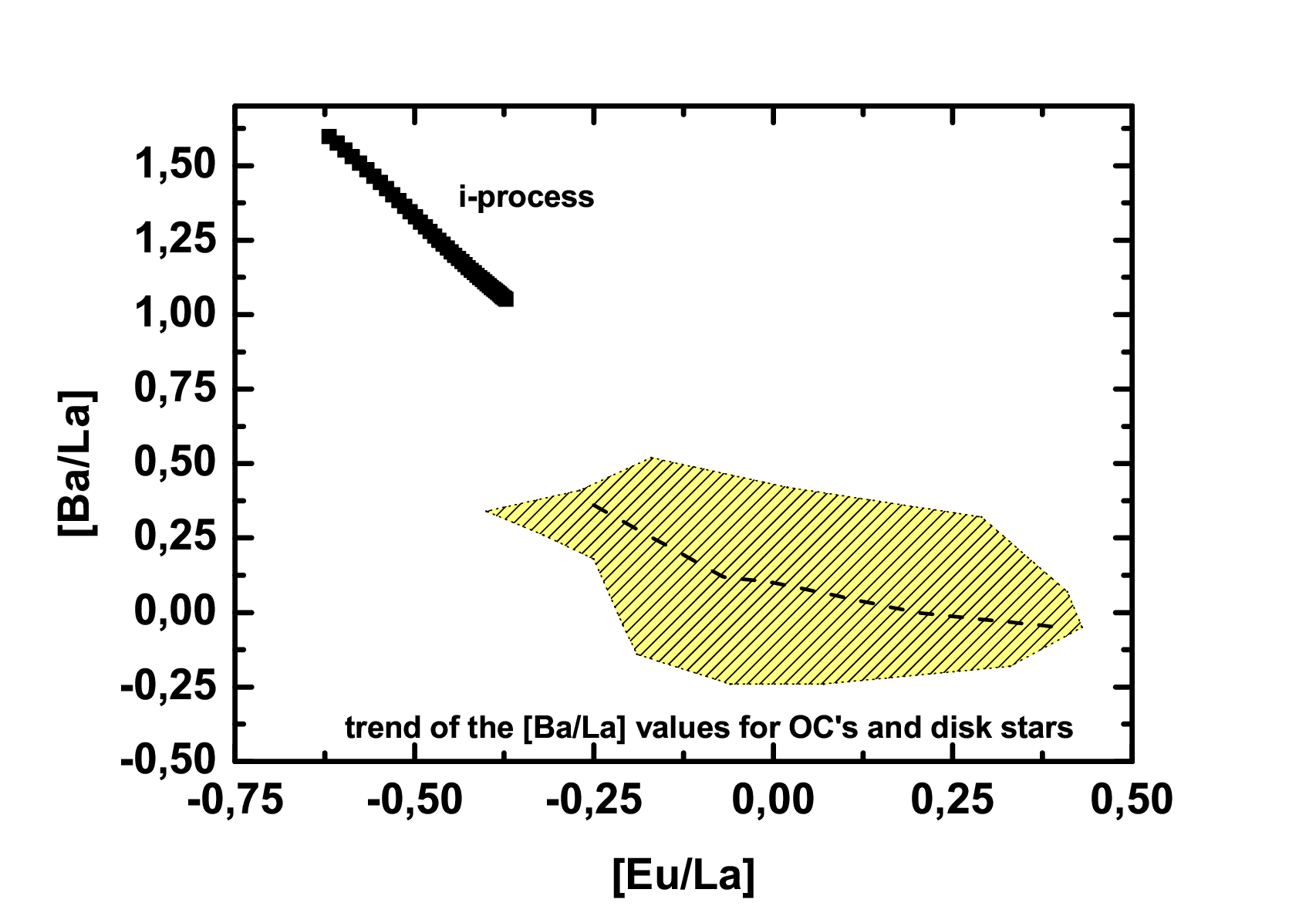}}
\caption[]{The [Ba/La] trend with respect to the [Eu/La] is shown for
the $i$-process trajectory  \citep{bertolli:13}, and
for the average of the OCs in the sample considered in this work.
The schematic observational distribution for OCs and disk stars
is shown.
Concerning the $i$-process
trajectory,
the earlier production of Ba compared to La is given by the radiogenic
contribution from $^{135}$I to $^{135}$Ba. With the increasing of the
total amount of neutrons, also La starts to be made and the [Ba/La]
tends to decrease. 
}
\label{iprocess}
\end{figure}

\begin{figure}
\begin{tabular}{c}
\includegraphics[width=8cm]{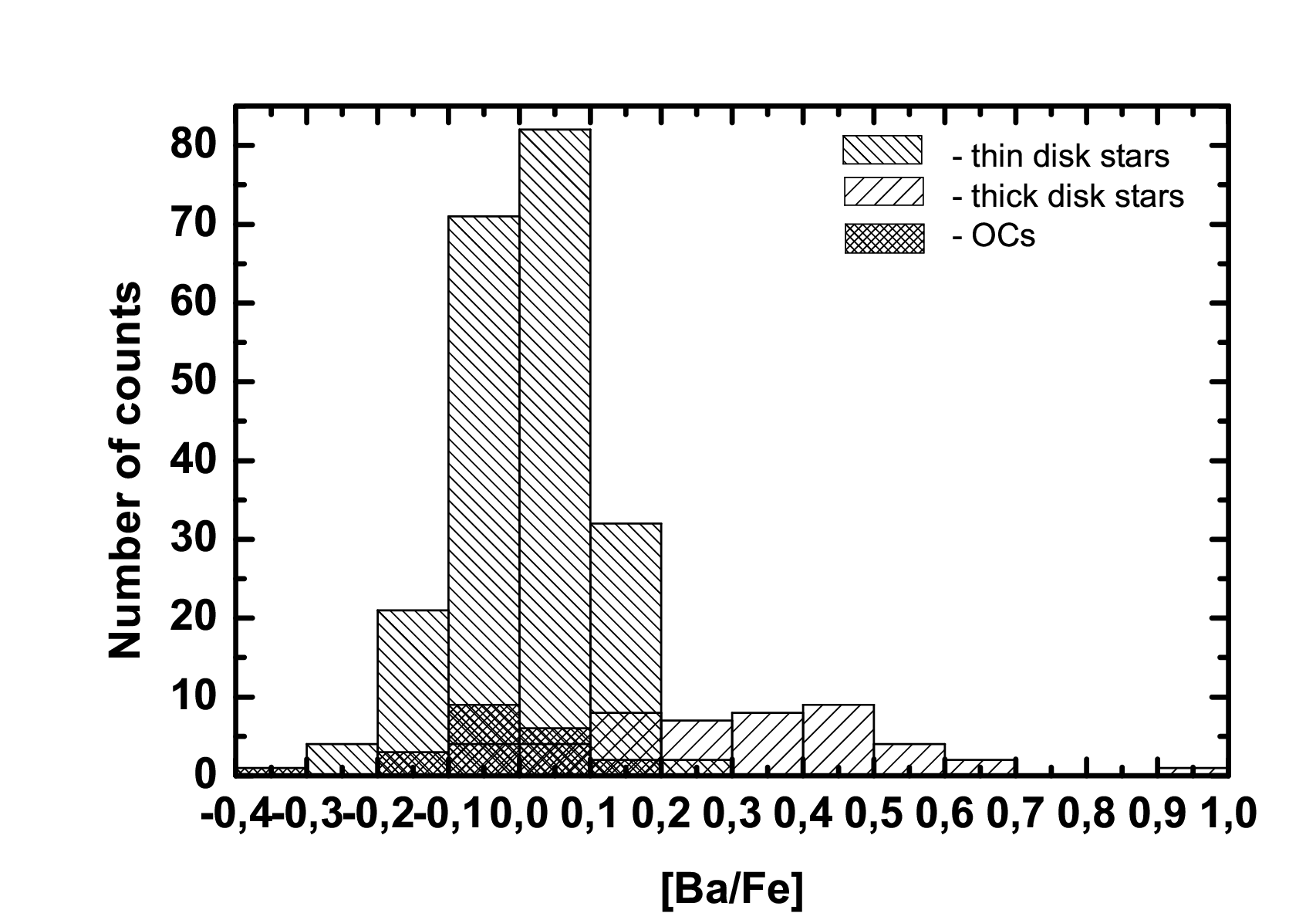}\\ \includegraphics[width=8cm]{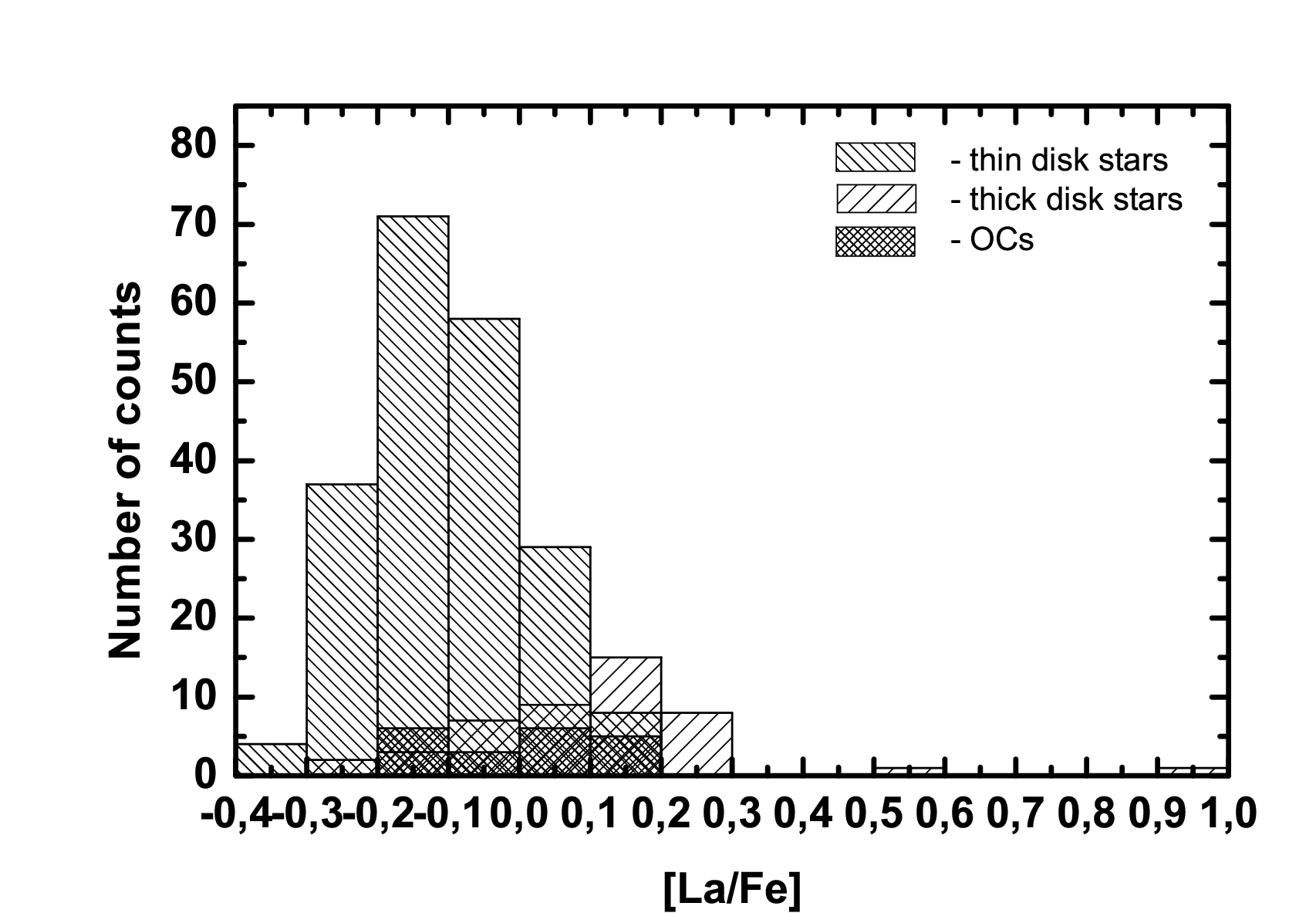}\\ \includegraphics[width=8cm]{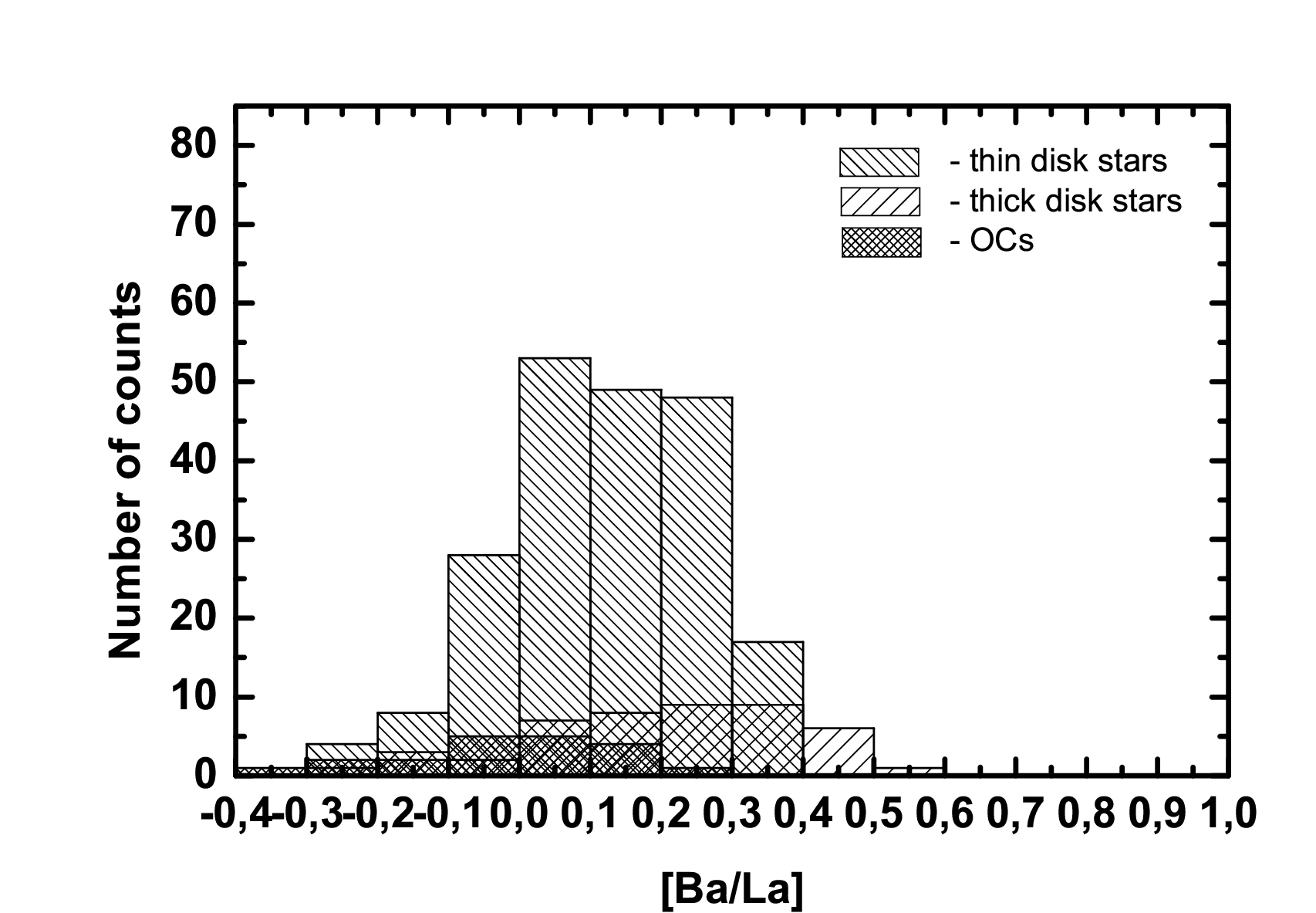}\\
\end{tabular}
\caption{The distribution of the [Ba/Fe], [La/Fe] and [Ba/La] ratio for OCs
is shown compared to  the thin disk stars and thick disk stars in
 Mishenina et al. (2013a).
}
\label{histo}
\end{figure}

Within the errors, we cannot observe any specific trend with
Age for Y and La. This result is consistent with previous
works, e.g., Jacobson \& Friel (2013). Yong et al. (2012) discussed
about a possible increasing trend of La with the Age of OCs,
but they also stressed about the large uncertainties and their
potential impact on those results. On the other hand, we confirm
the increasing average trend of [Ba/Fe] for younger OCs, in agreement
with previous works.

OCs measured by two independent groups with differences larger
than the error reported in Table~\ref{errors}, are connected
with a line. In the case when the values obtained by different
authors lie within the errors, in the figures we only report
the values obtained in this work, \cite{mis13b}, \cite{dor09}, \cite{dor12} and \cite{mai11}.
A remarkable case is NGC2141, where differences
between \cite{yon05} and \cite{jfr13} are about 0.5 dex
and 0.6 dex for [Ba/Fe] and [La/Fe], respectively.
These large differences are due in part from an average done
by \cite{jfr13} between two stars, 1007 and 1348, with 1007
showing a moderate Ba enrichment and a negative [La/Fe].
On the other hand, considering only the star 1348 in
common between the two authors, \cite{yon05} reported
[Ba/Fe] and [La/Fe] that are 0.4 dex and 0.44 dex
larger than \cite{jfr13}.
 While the differences affecting the Ba determination are already quite large
but they can be understood (according to their Table 3, \cite{jfr13} used the Ba lines
5853, 6141, and 6496 \AA\AA, with significant differences in the
resulting Ba abundances, while the abundance obtained from
the line 5853 \AA~in common with \cite{yon05} shows a better agreement),
we find more difficult to explain this discrepancy for La.

Despite these large differences, the conclusions
concerning the nature of the neutron-capture nucleosynthesis
signature in NGC~2141 will not
change considering \cite{yon05} or \cite{jfr13} observations.

As for  [Ba/Fe], three other OCs showing a significant departure are: 
NGC 752 -- [Ba/Fe] = 0.13 \citep{red12} and [Ba/Fe] = 0.52 \citep{car11};
Be 32 -- [Ba/Fe] = 0.22 \citep{jfr13} and [Ba/Fe] = 0.51
\citep{car11};
 NGC 2477 -- [Ba/Fe]  = 0.18  (this work) and [Ba/Fe]  = 0.48 \citep{bra08}.
Concerning La, large departures are present between Yong et al. (2012) and
Jacobson \& Friel (2013) for most of the common OCs, with the first authors
obtaining a larger La abundance.

In Figure~\ref{el_time}, the increasing spread of Ba enrichment
toward younger OCs and on average the much stronger enrichment of Ba compared
to La is confirmed within this larger sample of OCs.
This is difficult to explain in term of neutron-capture nucleosynthesis.
The production of Ba and La by neutron-capture processes is similar.
Ba and La are mostly made by their stable isotopes $^{138}$Ba and $^{139}$La,
located at the neutron shell closure N=82. They are commonly indicated
as $s$-process elements, since most of their abundance in the solar
system is explained by the $s$-process in Asymptotic Giant Branch stars.
In particular, according to Galactical Chemical Evolution simulations
about 85.2\% and 75.5\% of solar Ba and La
are made by the $s$-process \citep[][]{bisterzo:14}. Applying the residual
method where
the solar abundance of heavy elements beyond Fe is given by the contribution
of the $s$-process and the $r$-process \citep[e.g.,][]{arlandini:99},
the fraction of Ba and La made by the $r$-process are 14.8\% and 24.5\%
respectively. Therefore, the fact that Ba and La seem to have a different
behavior is puzzling. \cite{maiorca:12} proposed that the heavy elements
enrichment observed in young OCs is a signature of a larger $s$-process
enrichment from low mass AGB stars compared to the solar system,
and \cite{trippella:14} explored the impact of magnetic buoyancy
as a mechanism to trigger more efficient $s$-process production,
allowing to form more extended radiative $^{13}$C-pockets.
Nevertheless, an additional $s$-process contribution should not cause anomalies for the Ba/La ratio compared to established $s$-process calculations.
A larger enrichment of Ba compared to La it
is difficult to reconcile with $s$-process and
 $r$-process nucleosynthesis, or with a different combination of these
two components compared to the solar system.
To better explain this point, in Fig.~\ref{bala_eula} we show the [Ba/La]
compared to [La/Eu] for the OCs and the disk stars. The pure $s$-process
and $r$-process ratios are shown for comparison from \cite{travaglio:04}
and \cite{bisterzo:14}.

Within the scenario where the heavy elements are made by a combination of these two processes, the observations should fall inside the box, drawn by the assumptions of
pure $s$-process or $r$-process contributions. This is not the case for a sample of disk stars, and in particular for the most Ba-rich OCs.
From this figure it is clear that an additional $s$-process component cannot be the explanation of these anomalous abundances, since in this scenario the observations would still plot between the
solar system and the pure $s$-process lines.
In Fig.~\ref{bala_eula} we also show the [Ba/La] with respect to
the Age of the OCs. While there could be a mild increase of the
[Ba/La] ratio toward younger OCs, within the present uncertainties
we cannot derive any clear conclusion.

Three possible solutions to this puzzle are the following: 1) the measured Ba abundance for Ba-rich OCs (and part of the disk stars) is overestimated; 2) the measured La abundance is underestimated; 3) an additional
neutron capture component different from the $s$-process and the $r$-process
is contributing to the economy of heavy elements, producing more efficiently Ba than La and Eu.\\
Concerning the first option, we have discussed the possible sources of uncertainty affecting the estimation of the Ba abundance in Section~5. Here we just remind that we have considered NLTE effect for our analysis. Overall, we estimated that the uncertainty for Ba should not exceed 0.2 dex (Table \ref{errors}). On the other hand, for the [Ba/Fe] ratio 
in the literature we found discrepancies with our measurements in the order of 0.3 dex,
and in one case of 0.4 dex (see also the discussion in Jacobson \& Friel 2013,
and references therein). While we cannot discard this first option,
at the moment we would consider quite low the probability that this
is the solution of the Ba puzzle.
Indeed, among the all considered uncertainties the NLTE effect is the
only one that could explain a systematic overproduction for Ba, while
other uncertainties may also yield its underestimation from a given measure.
In the measurements reported in this work, we keep into account the
NLTE effect. In Section~4, we also discussed the possible issues
reported by \cite{dor12}. 

Concerning the second option, the uncertainty affecting the estimation of
the La abundance is lower than Ba, in the order of 0.1 dex (Table \ref{errors}).
The La lines adopted for \cite{mis14} and this work are 6320.41, 6390.48 \AA\AA.
There are not blending from other lines. The La abundance was
found taking into account the hyperfine structure.
The structure of electronic levels of La is similar to the structure of
the ones of Eu and as
\cite{mash00} has shown the NLTE corrections are very small for atoms of
europium. Therefore, we believe that the NLTE effects in lanthanum
abundance are also insignificant.

On the other hand, according to Jacobson \& Friel (2013) the use of the
EW's leads to reduce the La values by 0.07 dex, with an error
of $\pm$0.15 dex.
This last option would partly reduce the [Ba/La] ratio, but it cannot
explain the highest values shown in Fig. \ref{bala_eula}. We would
consider this last possibility alone unlikely to solve the Ba puzzle,
but we need to keep in mind the large differences obtained for
the La abundance between different authors mentioned earlier, beyond
the observational error.

If the first option is correct, the present observations in OCs would be easier to reconcile with GCE calculations using baseline AGB models and more in general with the prediction from the residual method. 
If the second option is correct, and the Eu observations are confirmed
compared to Fe, in order to explain the heavy element abundances in OCs a stronger $s$-process contribution may be needed, compared to
the solar system. As we mentioned before, a solution has been proposed laying in the present uncertainties on the physics mechanisms responsible for the formation of the $^{13}$C-pocket in AGB stars.
We could also argue that in order to reconcile the large observed
[Ba/La] with theoretical $s$-process predictions, both the first and the
second option are correct. In summary, at
the moment it is not clear how to explain a [Ba/La] up to 0.35 dex
sistematically higher than the pure $s$-process theoretical value, and
major observational issues would be required.

In case observations of Ba and La are correct, then this may be the first evidence of an
additional neutron-capture process contributing to the GCE of heavy elements. 
From Fig. \ref{bala_eula}, nor the $s$-process or the $r$-process can explain
a [Ba/La] larger than $\sim$ 0.15 dex.
In this case, the scenario that we propose is the additional contribution
from the intermediate neutron capture process, or $i$-process.
Firstly introduced more than thirty years ago by \cite{cowan:77}, the
$i$-process is characterized by neutron densities in the order of
10$^{15}$ neutrons cm$^{-3}$. As discussed by \cite{cowan:77},
the $i$-process is triggered by the mixing or ingestion of
H in He-burning stellar layers: protons are captured by the abundant He-burning product $^{12}$C forming $^{13}$C via the channel $^{12}$C(p,$\gamma$)$^{13}$N($\beta+$)$^{13}$C. $^{13}$C is the main source of neutrons via the ($\alpha$,n) reaction rate.

The first observational evidence of $i$-process activation in stars
is in the post-AGB Sakurai's object, explaining the anomalous heavy
element abundances \citep[][]{herwig:11} and the fast change of
abundance observed in a short timescale \citep[][]{asplund:99}.
Additional signature of the $i$-process activation in post-AGB stars
is found in presolar grains \citep[][]{jadhaw:13}.

Recently, \cite{bertolli:13} proposed the $i$-process as the source
of the anomalous heavy-element signature observed in a sub-sample
of Carbon-enhanced Metal Poor stars, usually explained as a mixture
of $s$-process and $r$-process contribution
\citep[CEMP-rs stars, e.g.,][]{masseron:10,lugaro:12,bisterzo:14}.
In particular, some of these stars seem to show a [Ba/La] ratio
larger than what the $s$-process, the $r$-process or a combination of
them can explain.

The $i$-process can potentially explain this larger ratio,
where the bulk of Ba is radiogenic, made by the decay of $^{135}$I, but major nuclear uncertainties still affect thereotical simulations \citep[][]{bertolli:13}.

Another major problem is that multi-dimensional hydrodynamics simulations are needed in order to produce
consistent results for the H ingestion in He-burning layers
\citep[e.g.,][]{herwig:07,mocak:11,stancliffe:11,herwig:13,woodward:13}.
Baseline hydrostatic stellar models
can provide only qualitative information at best about these events,
and without guidance from hydrodynamics simulations fail to reproduce
the observations \citep{herwig:11}.

In Fig.~\ref{iprocess} we show the theoretical $i$-process
predictions from a simple trajectory reproducing the $i$-process
neutron density conditions.
This is the same trajectory given by \cite{bertolli:13},
but adopting initial abundances beyond Si compatible with
galactic disk metallicity. In particular, the initial $^{56}$Fe
mass fraction is 5.3$\times$10$^{-4}$, about half of the $^{56}$Fe
amount in the Solar System.

In the figure, the trajectory behavior is shown when [Ba/Y] is larger
than 0.7 dex (consistent with observations of Ba-rich OCs), and for Ba
five times larger than Pb. The trajectory shows [Ba/La] high values,
decreasing with the increasing of the amount of  neutrons available.
For comparison, we also report the average trend observed for all the
OCs, and the observed range covered by OCs and disk stars considered
here (see Fig.~\ref{bala_eula}). 
From Fig.~\ref{iprocess}, we can argue that a combination of the
$i$-process, the $s$-process and the $r$-process provide a scenario
capable to explain the observed large [Ba/La], mainly contributing
to Ba compared to Y, La and Eu.

At the moment it is difficult to constrain what is(are)
the host(s) of the $i$-process.
If this scenario is correct, the $i$-process occurrence cannot be
limited to H-deficient post-AGB stars \citep{herwig:11}, or
to low metallicity stellar hosts \citep{bertolli:13}.
For instance, in Pignatari et al. (2014, in prep.) we found proof of late
H ingestion in massive stars just before the CCSNe explosion. The fact that
these H-ingestion  events in massive stars
are also associated with $i$-process production is not clear, and need more
investigation. 

In Fig.~\ref{bala_eula}, we show that also a relevant fraction of the disk
stars show a [Ba/La]
larger than the $s$-process limit. In this case, the departure is lower
compared to the most Ba-rich OCs, and on average at larger [Eu/La]. 
As also shown from previous works for OCs and for disk stars
\citep[e.g.,][]{dor09,bensby05,jfr13}, the average
Ba enrichment seems to increase for objects younger than the Sun.
This could suggest that in these last Gyrs the $i$-process contribution
is becoming more relevant than what it was for our Sun, compared to the
established $s$-process and $r$-process contribution.
On the other hand, as we mentioned earlier we cannot claim any trend of
the [Ba/La] with respect to the Age of the OCs (Fig.~\ref{bala_eula}).

Taking into account the previous discussion about observation uncertanties,
it is not possible to derive any strong conclusion,
claiming that also disk stars are hiding an $i$-process contribution.
On the other hand, in case OCs are carrier of an $i$-process component,
it would be reasonable to assume that the same is happening for disk stars. 

In Fig. \ref{histo}, we show the [Ba/Fe], [La/Fe] and [Ba/La] distribution
for OCs compared to the thin disk stars by \cite{mis13a}. Concerning the
[La/Fe] distribution, OCs show a peak shifted by 0.1--0.2 dex compared to
thin disk stars. While this difference is not negligible,
it is still within the present observational uncertainties. 

On the other hand, the [Ba/Fe] in OCs is clearly shifted toward larger values,
and the distribution looks more scattered than for disk stars.
Note that the present observed scatter is real, represented in
our sample of OCs \citep[][and this work]{mis13b}, and not a product of observational
systematics uncertainties.
But concerning the analysis including all the OCs, in agreement with previous
works we confirm that the observational uncertainties and the lack of
homogeneous abundance analysis is an issue that needs to solved in the
future in order to derive definitive conclusions.
Finally, the larger [Ba/Fe] spread in OCs compared to disk stars is conserved in the [Ba/La] ratio. The disk stars show a distribution peak 0.1 dex larger than solar, consistent within the uncertainties. For OCs the peak is much broader, and shifted to larger values.
The reason of these differences in the heavy element enrichment between OCs and disk stars has to be analyzed with chemical evolution simulations, and it is not the goal of this paper. According to Fig.~\ref{bala_eula}, a possible scenario {to explain the observed increase of [Ba/La] with the decreasing of [Eu/La]} could be that OCs have
overall a smaller contribution from the $r$-process compared to disk stars,
highlighting  contribution from $s$-process and $i$-process components.

In Fig.~\ref{el_fe} and~\ref{el_time}, we highlighted the OCs
Be~31 and NGC~2141 by \cite{yon05}, showing a [La/Fe] much larger than other OCs (see also Fig.~\ref{histo}).
Be~31 is one of the most metal poor OCs presently known, despite it is not one of the oldest (about 2 Gyr, \cite{car94}).
It shows a much larger $s$-process enrichment of $s$-process elements Ba and La
compared to the $r$-process element Eu. \cite{yon05} explained this as an
affect of a significant contribution from AGB stars $s$-process rich material.
The position of Be~31 in Fig.~\ref{bala_eula} seems to confirm this scenario.
The [Eu/La] is consistent with a larger $s$-process contribution compared
to the Sun, while the [Ba/La] can be explained by an enrichment history
given by the $s$-process and the $r$-process contributions.
Within this scenario, due to the relative low metallicity we would expect
that Be~31 has also a larger Pb enrichment compared to the Sun.
Therefore, the measurement of Pb would be extremely useful to
confirm this scenario.
Unfortunately, at the moment there are no available measurements for Pb
abundances in OCs.
NGC~2141 has an Age similar to Be~31, but has a metallicity much closer
to the Sun. It  may be the most Ba-rich OC within the sample considered
in this work  but we have shown that large discrepancies are obtained
considering different works (e.g., Fig.~\ref{el_time} and
Table~\ref{comparison_literature}). Within the large uncertainties
affecting [Ba/Fe] and [La/Fe],
the [Ba/La] is larger than the $s$-process ratio, consistently with the
anomalous signature discussed in this work.
Assuming that there are no other observational issues for Ba, NGC~2141
is another candidate where the $i$-process contribution can be identified.

\section{Conclusions and final remarks}

In this work we presented and discussed new abundance measurements for five OCs:
Cr~110, Cr~261, NGC~2477, NGC~2506 and NGC~5822. We analyzed these new
results for neutron-capture elements 
complementing them with literature data.
Literature data show significant author-to-author differences, that we discussed.
Beside these differences, we found confirmation of the larger
scatter of the Ba abundance in OCs compared to disk stars. 
We also confirm that the average Ba abundance is increasing for younger OCs, while there is not clear trend with the metallicity [Fe/H].
To a lower extent, the [La/Fe] ratio seems to show a similar behavior as Ba. 

With the exception of few OCs, the [La/Fe] is found to be consistent with
the average disk enrichment. 
A possible source of uncertainty is, however,  the impact of the metallicity dependence of the Fe
production from SNIa. This needs future investigations because of its implications for the ratio between
neutron-capture elements and Fe in thin-disk stars  and  OCs.
Finally, the [Y/Fe] ratio in OCs is consistent with disk stars and the
Sun within the uncertainties.

Besides the overall enrichment of neutron-capture elements  compared
to Fe, we showed that the resulting [Ba/La] ratio is not consistent
with the established scenario for the production of heavy elements,
with a combined contribution from the $s$-process and $r$-process only.

 The OCs (and to less extent disk stars) show an increasing [Ba/La]
ratio for a decreasing [Eu/La], reaching values for [Ba/La] up to
0.4--0.5 dex, 
much larger than what expected from $s$-process or the $r$-process.

We considered three main options to explain this occurrence.

The first two options are related to possible observational issues with
Ba and La.
In particular, we discuss possible uncertainties affecting the measurement
of the Ba abundance.
We argue that the uncertainty in the Ba abundance alone cannot explain
the enrichment observed for such  a large number of OCs. On the other
hand, it might be possible that the La abundance be under-estimated,
which  could help to reconcile the OCs observations within
the baseline scenario of an $s$-process and $r$-process contribution. 
Therefore, the reduction of present observational uncertainties and of
the amount of inhomogeneity of the observed data could still solve this puzzle.

In case instead the observations are correct, we considered and
discussed a third option: that  OCs are  showing the additional
contribution from the $i$-process. 
This may be the first evidence that the $i$-process had a relevant
contribution to the galactic chemical evolution of the Galaxy,
together with the $s$-process and the $r$-process. 
One of the peculiar signatures of the $i$-process is to predict a [Ba/La] ratio much
larger than the $s$-process or the $r$-process, within the observed spread
of [Eu/La]. 
The capability to disentangle the production of Ba and La is a unique feature,
caused by neutron densities
intermediate between the $s$-process and the $r$-process. We show that the
additional contribution from the $i$-process is consistent with the present
observations in OCs.

This scenario needs to be corroborated by considering more neutron-capture
elements.
Despite the fact that the $i$-process was defined more than 30 years
ago, only in the last five years we are starting to collect
observational evidences of its existence 
in different types of stars at different metallicities, and in presolar grains.
Furthermore, robust predictions of $i$-process stellar yields cannot be provided by baseline one-dimensional hydrostatic models. The $i$-process is associated to H ingestion in hot He-burning environments, requiring the guidance of multi-dimensional hydrodynamics simulations. These are challenging and computationally expensive but feasible, as proven from a number of simulations that are becoming available. The impact of present nuclear uncertainties on the $i$-process nucleosynthesis should also be fully explored, if relevant for the observed elemental ratios.
We refer to \cite{bertolli:13} for the nuclear uncertainties affecting the [Ba/La] ratio.

In conclusion, the present work provided an important step forward for  our understanding of the nucleosynthesis of neutron-capture elements in the Galaxy.
More observations are needed for more neutron capture elements in OCs. On the other hand, when discrepances larger than about 0.2 dex exist between different works for the same objects, a new independent analysis would be recommended. It is obvious that the inhomogeneity of the data we gathered from
the literature  is playing a  major obstacle toward a solid 
understanding of these abundance ratio patterns. Unfortunately this is
the actual situation, and we will need to wait for more extended and
homogeneous data-set to become avalailable, such as Gaia-ESO
\citep[][]{gilmore12}. Similar conclusions have been derived from other previous
works analyzing the abundances for an extended sample of OCs.
This consensus is an important step to draw the priorities for next
observational campaigns.

On the theoretical side, the calculation of robust $i$-process yields
for different types of stars are not available at the moment.
This will require extensive hydrodynamics simulations for different
stellar environments in the coming years, as a guidance for complete
sets of one-dimensional hydrostatic stellar models.

\section*{Acknowledgements}
We thank the anonymous referee for a detailed and thoughtful review of our
manuscript.
T. Mishenina, S. Korotin and M. Pignatari thank for the support from the Swiss National Science Foundation,
project SCOPES No. IZ73Z0$_{}$152485.
MP and FH acknowledge significant support to NuGrid from NSF grants PHY 02-16783 and PHY 09-22648 (Joint Institute for Nuclear Astrophysics, JINA) and EU MIRG-CT-2006-046520.
M.~P.\ acknowledges an Ambizione grant of the SNSF and support from SNF (Switzerland). F.~H.\ acknowledges NSERC Discovery Grant funding.

\label{lastpage}

\bsp

\end{document}